\documentclass[12pt,letter]{article}
\usepackage{setspace}
\usepackage{amsmath,amssymb,graphics,epsfig,latexsym,color}
\usepackage{times}
\usepackage{subfigure}
\usepackage{psfrag}
\usepackage{graphicx}

\newcommand{\increaseUDLR}[4]{
        \addtolength{\textwidth}{#3}
        \addtolength{\textwidth}{#4}
        \addtolength{\textheight}{#1}
        \addtolength{\textheight}{#2}
        \addtolength{\oddsidemargin}{-#3}
        \addtolength{\evensidemargin}{-#3}
        \addtolength{\topmargin}{-#1}
}
\increaseUDLR{20mm}{20mm}{20mm}{20mm}

\newcommand{\qed}{\hfill  \rule{2mm}{3mm}}

\newtheorem{lemma}{Lemma}
\newtheorem{theorem}[lemma]{Theorem}

\newtheorem{definition}[lemma]{Definition}

\newcommand{\junk}[1]{}
 
\newcommand{\argmin}{\operatornamewithlimits{argmin}}
\newcommand{\argmax}{\operatornamewithlimits{argmax}}

\newenvironment{proof}
{\smallskip\noindent{\bf Proof.\hspace{1ex}}}{\par\bigskip}

\newcommand{\abf}{\mbox{${\bf a }$} }
\newcommand{\bbf}{\mbox{${\bf b }$} }
\newcommand{\cbf}{\mbox{${\bf c }$} }

\newcommand{\gbf}{\mbox{${\bf g }$} }

\newcommand{\ubf}{\mbox{${\bf u }$} }

\newcommand{\xbf}{\mbox{${\bf x }$} }

\newcommand{\Abf}{\mbox{${\bf A }$} }
\newcommand{\Bbf}{\mbox{${\bf B }$} }
\newcommand{\Cbf}{\mbox{${\bf C }$} }

\newcommand{\Gbf}{\mbox{${\bf G }$} }

\newcommand{\Kbf}{\mbox{${\bf K }$} }
\newcommand{\Mbf}{\mbox{${\bf M }$} }

\newcommand{\Pbf}{\mbox{${\bf P }$} }

\newcommand{\Ubf}{\mbox{${\bf U }$} }

\newcommand{\Xbf}{\mbox{${\bf X }$} }

\newcommand{\ZZ}{\mathbb Z}

\newif{\ifbibtex}
\bibtexfalse

\newcommand{\field}{2}
\newcommand{\base}{\mathbb{F}_{\field}}
\newcommand{\ext}{\mathbb{F}_{\field^T}}

\newcommand{\tr}{{\rm Tr}_{\field^T/\field}}
\newcommand{\Complex}{{\rm {\bf C}\mkern-9mu\rule{0.05em}{1.4ex}\mkern10mu}}
\newcommand{\g}[1]{\mathbf{g}^{(#1)}}
\newcommand{\tg}[1]{\tilde{\mathbf{g}}^{(#1)}}
\newcommand{\hg}[1]{\hat{\mathbf{g}}^{(#1)}}
\newcommand{\bg}[1]{\breve{\mathbf{g}}^{(#1)}}

\title{Embedded Rank Distance Codes for ISI channels}

\author{
S. Dusad\thanks{
EPFL, Lausanne, Switzerland, 
    Email: \{sanket.dusad,suhas.diggavi\}@epfl.ch. 
S. Dusad was supported in part by SNSF Grant \# 200021-105640/1. S
N. Diggavi is part of the SNF NCCR-MICS center on wireless sensor
networks.  }
\hspace{0.2in} S. N. Diggavi\footnotemark[1]
\hspace{0.2in} A. R. Calderbank\thanks{Princeton University, Email: 
         calderbank@math.princeton.edu. A R. Calderbank was supported in part by
NSF grant \# 1096066.} }

\begin{document}
\maketitle

\begin{abstract}
Designs for transmit alphabet constrained space-time codes naturally
lead to questions about the design of rank distance codes. Recently,
diversity embedded multi-level space-time codes for flat fading
channels have been designed from sets of binary matrices with rank
distance guarantees over the binary field by mapping them onto QAM and
PSK constellations. In this paper we demonstrate that diversity
embedded space-time codes for fading Inter-Symbol Interference (ISI)
channels can be designed with provable rank distance guarantees. As a
corollary we obtain an asymptotic characterization of the fixed
transmit alphabet rate-diversity trade-off for multiple antenna fading
ISI channels. The key idea is to construct and analyze properties of
binary matrices with a particular structure induced by ISI channels.

\end{abstract}

\section{Introduction}
\label{sec:intro}
Over the past decade significant progress has been made in
constructing space-time codes that achieve the optimal rate-diversity
trade-off for {\em flat-fading} channels when there are transmit
alphabet constraints \cite{TSC98,LuKumar05}. Far less attention has
been given to space-time code design and analysis for fading channels
with memory, {\em i.e.,} Inter-Symbol Interference (ISI) channels
which are encountered in broadband multiple antenna
communications. There have been several constructions of space-time
codes for fading ISI channels using multi-carrier techniques (see for
example \cite{SSOL03} and references therein). However, since these
inherently increase the transmit alphabet size, and the right
framework to study such constructions is through the {\em
diversity-multiplexing} trade-off \cite{TseViswanath05}. We examined
diversity embedded codes for ISI channels in \cite{DD06}, by
considering the diversity-multiplexing trade-off.

As in space-time code design for flat-fading channels, it is natural
to ask for a characterization of the rate-diversity trade-off for ISI
channels with transmit alphabet constraints\footnote{Throughout this
paper we restrict our attention to a transmit alphabet constraint,
{\em i.e.,} the transmit alphabet is restricted to be from the set
$\mathcal{A}$. Therefore this imposes a maximal rate of
$M_t\log|\mathcal{A}|$ bits and we normalize the rate by
$\log|\mathcal{A}|$ and state the rate in terms of a number in
$[0,M_t]$ symbols per transmission.}. The problem of constructing
space-time codes with fixed transmit alphabet constraints is partially
motivated by the need to control the transmit spectrum as well as the
peak-to-average (PAR) ratio of the transmitted signal. For example, if
we restrict transmission to PSK alphabet, it is clear that we have a
unit peak-to-average ratio (PAR) making it possible to use efficient
non-linear amplifiers (requiring small PAR), which are more efficient
and hence suitable for mobile devices. Another important reason to
consider this problem is a fundamental theoretical question, which is
motivated by the origins of space-time codes for flat-fading channels
in \cite{TSC98} where the constructions were for fixed transmit
alphabet. For this constraint, there exists a trade-off between rate
and diversity, for the flat-fading case.  In this paper we ask the
corresponding question for fading ISI channels. Since space-time code
design with diversity order guarantees requires control over the rank
distance of the codewords \cite{TSC98}, the main topic of this paper
is to design codes with rank distance guarantees for ISI channels.

Diversity embedded codes were introduced in \cite{Diggavi:DIMACS}
which allowed different parts of a message to have different diversity
order guarantees. These codes allowed diversity to be viewed as a
systems resource that can be allocated judiciously to achieve a
target rate-diversity trade-off in wireless communications.  A
class of such multi-level diversity embedded codes suitable for
flat-fading channels was constructed in \cite{DDCA05,CDA04,DCDA06}. In
this paper we extend these constructions to ISI channels. 

The corresponding question, of what is studied in this paper, can be
also be posed in the context of the trade-off between diversity and
multiplexing rate. Such an information-theoretic question, for the
flat-fading case, has been posed and partially answered in
\cite{DigTse05,DigTse06}. For scalar ISI channels, we have studied
code designs for rate-growth (multiplexing rate) codes and the
diversity embedding properties in \cite{DD06}. There we have shown
that the diversity multiplexing trade-off for the scalar ISI channel
is actually successively refinable. However, the code designs and
criteria for the rate-growth codes are quite different from those
needed for the fixed rate, transmit alphabet constrained codes, which
are the focus of this paper.

For the case of a scalar ISI channel with $\nu+1$ taps and a single
transmit antenna, it can be shown by a simple argument (see for
example \cite{TseViswanath05}) that an uncoded transmission scheme can
achieve a diversity order of $(\nu+1)$. The best case scenario for the
rate-diversity trade-off for ISI channels with multiple transmit
antennas would be similar to the flat-fading case, but with a
$(\nu+1)-$fold increase in the diversity order.  However, in the
multiple transmit antenna case, it is not obvious that a space-time
code designed for a flat-fading channel can achieve such a
$(\nu+1)-$fold increase in the diversity order. All that can be
guaranteed is that a space-time code that achieves diversity order $d$
over a flat-fading channel will still achieve diversity order $d$ over
a fading ISI channel \cite{TSC98}. In particular in Example 1 of
Section \ref{sec:examples} we provide an example of a code which
achieves particular points on rate-diversity trade-off for flat-fading
channels and fails to do so in the case of ISI channels. Therefore,
the design of codes for fading ISI channels cannot be immediately done
by using the codes for flat-fading channels. However, in this paper we
see that codes designed for the fading ISI channel can be used
successfully to achieve the rate-diversity trade-off for the
flat-fading case as well.

A finite alphabet construction to exploit the potential diversity gain
from ISI channels with $M_t$ multiple transmit antennas was proposed
in \cite{GHLF03} for the maximal diversity case. But the rate of the
code for this construction was $1/M_t$ as opposed to the maximal
potential rate of $1$. In this paper we show that as the transmission
block size increases we can construct codes that have rate $1$ and
achieve the maximal diversity order of $(\nu+1)M_t$.  We characterize
the rate diversity tradeoff for the ISI channels and construct codes
which achieve this tradeoff (asymptotically in block size). We build
on the construction technique introduced in \cite{DCDA06} to design
diversity embedded codes for ISI channels that guarantee multiple
reliability (diversity) levels. Given that we can achieve a
$(\nu+1)-$fold increase in the diversity order for ISI channels, this
flexibility could be quite important.

The main contributions of this paper are as follows. We extend the
rate-diversity trade-off bound from \cite{TSC98} and develop the
diversity embedded code design criteria for fading ISI channels in
Section \ref{sec:ModelCD}. The basic multi-level construction of
diversity-embedded space-time code for fading ISI channels is given in
Section \ref{sec:MLcons}. We also show that this construction can be
specialized to asymptotically achieve the diversity-rate trade-off for
ISI channels. The key ingredient is the construction of binary codes
for ISI channels with rank-distance guarantees, and this is done in
Section \ref{sec:Rate} and Section \ref{sec:Rank}.  This is perhaps
the most important technical contribution of this paper.  We also
construct of convolutional codes suitable for transmission over the
ISI channel in Section \ref{sec:Trellis}.  In Section
\ref{sec:examples} we give examples of codes constructed by the method
given in the paper along with their numerical performance.

\section{Problem Statement and code design criteria}
\label{sec:ModelCD}
In Section \ref{subsec:ChanModel}, we define the ISI channel model .
Section \ref{subsec:STCode} recalls the code design criteria for
diversity embedded codes for flat-fading channels given in
\cite{DCDA06} and extends it to the fading ISI case. These criteria
give the connection between embedded rank-distance codes and
diversity-embedded space-time codes.  The rate-diversity trade-off for
flat-fading channels is reviewed in Section \ref{subsec:RateDiv}, and
a simple upper bound for the corresponding trade-off for the fading
ISI case is established. The subsections \ref{subsec:setpart} and
\ref{subsec:AlgProp} are based on \cite{DCDA06} and reproduced here
for completeness.  In Section \ref{subsec:setpart}, we review the
principle of set-partitioning and give algebraic properties of such
partitions in Section \ref{subsec:AlgProp}. These properties would be
useful in {\em lifting} rank properties of binary matrices over binary
fields to the complex domain, thereby giving provable diversity
embedded code constructions.

\subsection{Channel Model}
\label{subsec:ChanModel}
Our focus in this paper is on the quasi-static frequency selective
(ISI) channel with $(\nu+1)$ taps where we transmit information coded
over $M_t$ transmit antennas and have $M_r$ antennas at the
receiver. Furthermore, we make the standard assumption that the
transmitter has no channel state information, whereas the receiver is
able to perfectly track the channel. The code is designed over a large
enough block size $T\geq T_{thr}$ transmission symbols, where
$T_{thr}$ is specified in the constructions given in Section
\ref{sec:MLcons}. The received vector at time $n$ after demodulation
and sampling can be written as,
\begin{equation}
{\bf y}[n] = {\bf H}_0 {\bf x}[n] + {\bf H}_1 {\bf x}[n-1] +\ldots
+{\bf H}_{\nu}{\bf x}[n-\nu] + {\bf z}[n]
\end{equation}
where, ${\bf y} \in \Complex^{M_r\times 1}$, ${\bf H}_l \in
\mathbb{C}^{M_r \times M_t}$ represents the matrix ISI channel, ${\bf
x}[n] \in \Complex^{M_t\times 1}$ is the space-time coded transmission
sequence at time $n$ with transmit power constraint $P$ and ${\bf z}
\in \Complex^{M_t\times 1}$ is assumed to be additive white
(temporally and spatially) Gaussian noise with variance
$\sigma^2$. The matrix ${\bf H}_l$ consists of fading coefficients
$h_{ij}$ which are i.i.d. $\mathcal{C}\mathcal{N}(0,1)$ and fixed for
the duration of the block length ($T$).

Consider a transmission scheme in which we transmit over a period
$T-\nu$ and send (fixed) known symbols\footnote{Taken without loss of
generality to be $0$.} for the last $\nu$ transmissions. For the period of
communication we can equivalently write the received data as,
{\footnotesize
\begin{align}
\label{eqn:model1}
\underbrace{
\left[
\begin{array}{ccc}
{\bf y}[0] & \ldots & {\bf y}[T-1]
\end{array}
\right]
}_{\mathbf{Y}}
&=
\underbrace{
\left[
\begin{array}{ccc}
{\bf H}_0 & \ldots & {\bf H}_{\nu}
\end{array}
\right]
}_{\mathbf{H}}
\underbrace{
\left[
\begin{array}{ccccccc}
{\bf x}[0] & {\bf x}[1] & \ldots     & {\bf x}[T-\nu-1] & 0              & \ldots     & 0 \\
0          & {\bf x}[0] & {\bf x}[1] & \ldots         & {\bf x}[T-\nu-1] & 0          & 0 \\
\ldots     &            & \vdots     & \ddots         & .              & .          & \vdots\\
0          & \ldots     &  0         &{\bf x}[0]      & {\bf x}[1]     & \ldots     & {\bf x}[T-\nu-1]  
\end{array}
\right]
}_{\mathbf{X}} +{\bf Z} 
\end{align}
}
{\em i.e.},
\begin{align}
\label{eqn:model2}
{\bf Y} & = {\bf H} {\bf X} + {\bf Z} 
\end{align}
where ${\bf Y} \in \mathbb{C}^{M_r \times T } $, ${\bf H}\in
\mathbb{C}^{M_r \times (\nu+1) M_t }$, ${\bf X}\in \mathbb{C}^{(\nu+1)
M_t\times T }$, ${\bf Z}\in \mathbb{C}^{M_r \times T}$.  Notice that
the structure in (\ref{eqn:model1}) is different from the flat-fading
case, since the channel imposes a Toeplitz structure on the equivalent
space-time codewords $\mathbf{X}$ given in
(\ref{eqn:model1})-(\ref{eqn:model2}). This structure makes the design
of space-time codes different than in the flat-fading case. For reference, 
the space-time codeword is completely determined by the matrix ${\bf X}^{(1)}$
given by,
\begin{align}
\label{eq:X1def}
\mathbf{X}^{(1)} &=  \left[  
\begin{array}{ccccccc}
{\bf x}[0] & {\bf x}[1] & \ldots  & {\bf x}[T-\nu-1] & 0    & \ldots         & 0 
\end{array}
\right ]
\end{align}

\subsection{Diversity-embedded code design criteria}
\label{subsec:STCode}

A scheme with diversity order $d$ has an error probability at high SNR
behaving as $\bar{P}_e(\mbox{SNR}) \approx \mbox{SNR}^{-d}$
\cite{TSC98}. More formally,

\begin{definition}
\label{defn:div}
A coding scheme which has an average error probability
$\bar{P}_e(\mbox{SNR})$ as a function of $\mbox{SNR}$ that behaves as
\begin{equation}
\label{eq:DivDefn}
\lim_{\mbox{SNR}\rightarrow \infty} \frac{\log(\bar{P}_e(\mbox{SNR}))}{\log(\mbox{SNR})} = -d
\end{equation}
is said to have a diversity order of $d$.
\end{definition}

The fact that the diversity order of a space-time code is determined
by the rank of the codeword difference matrix is well known
\cite{TSC98,Guey99}. Therefore, for flat-fading channels, it has been
shown that the diversity order achieved by a space-time code is given
by \cite{TSC98}
\begin{equation}
\label{eq:DivTSC}
d = M_r\min_{\Cbf_1\neq\Cbf_2} \mbox{rank}(\Cbf_1-\Cbf_2) \; ,
\end{equation}
where $\Cbf_1,\Cbf_2\in\Complex^{M_t\times T}$ are the space-time
codewords. Clearly the analysis in \cite{TSC98,Guey99} can be easily
extended to fading ISI channels, and we can write
\begin{equation}
\label{eq:DivISI}
d = M_r\min_{\Xbf_1\neq\Xbf_2} \mbox{rank}(\Xbf_1-\Xbf_2) \; ,
\end{equation}
where $\Xbf_1,\Xbf_2\in\mathbb{C}^{(\nu+1) M_t\times T }$ are matrices
with structure given in (\ref{eqn:model1}).

It is easy to see from the structure of ${\bf X}$ in
(\ref{eqn:model1}) that the rank of the matrix ${\bf X}$ is {\em at
most} $(\nu+1)$ times the rank of the matrix ${\bf X}^{(1)}$ (see
(\ref{eq:X1def})), {\em i.e.},
\begin{align}
{\rm rank}(\mathbf{X}) & \leq (\nu+1){\rm  rank}({\bf X}^{(1)})  \label{upper_bound}
\end{align}

The codebook structure proposed in  \cite{DCDA06} takes two
information streams and outputs the transmitted sequence
$\{\xbf(k)\}$. The objective is to ensure that each information stream
gets the designed rate and diversity levels. Let ${\mathcal E}$ denote
the message set from the first information stream and ${\mathcal F}$
denote that from the second information stream. Then analogous to
Definition \ref{defn:div}, we can write the diversity order for the
messages as,
\begin{equation}
\label{eq:DivTuple}
D_a =\lim_{\mbox{SNR}\rightarrow\infty}
\frac{\log\bar{P}_e(\mathcal{E})}{\log(\mbox{SNR})} ,\,\, D_b
=\lim_{\mbox{SNR}\rightarrow\infty} \frac{\log\bar{P}_e(\mathcal{F})}{\log(\mbox{SNR})}.
\end{equation}
\paragraph*{Design criteria for fading ISI channels:} 
The space-time codeword for fading ISI channels have the structure
given in (\ref{eqn:model1}). To translate this to the diversity
embedded case, we annotate it with given messages $\abf\in {\mathcal
E},\bbf\in \mathcal{F}$, as $\Xbf_{\abf,\bbf}$. Clearly we can then
translate the code design criterion from (\ref{eq:DivISI}) to
diversity embedded codes for ISI channels as,
\begin{equation}
\label{eq:DivCnstA}
\min_{\abf_1\neq\abf_2\in {\mathcal E}}\min_{\bbf_1,\bbf_2\in {\mathcal F}}
\mbox{rank}(\Xbf_{\abf_1,\bbf_1}-\Xbf_{\abf_2,\bbf_2})) \geq D_a/M_r
\end{equation}
In an identical manner, we can show for the message set $\mathcal{F}$,
we need the following to hold.
\begin{eqnarray}
\label{eq:2LayerDes2}
\min_{\bbf_1\neq\bbf_2\in {\mathcal F}}\min_{\abf_1,\abf_2\in {\mathcal E}}
\mbox{rank}(\Xbf_{\abf_1,\bbf_1}-\Xbf_{\abf_2,\bbf_2})) \geq D_b/M_r.
\end{eqnarray}
As one can easily see, these are simple generalizations of the diversity-embedded
code design criteria developed in \cite{Diggavi:DIMACS} to the fading ISI case.

\subsection{Rate-Diversity Trade-off for Flat Fading Channels}
\label{subsec:RateDiv}

For a given diversity order, it is natural to ask for upper bounds on
achievable rate.  For a flat Rayleigh fading channel, this has been
examined in \cite{TSC98} where the following result was obtained.
\begin{theorem}
\label{thm:RateDivTSC}
( \cite{TSC98,LuKumar03}) Given a constellation of size
$|\mathcal{A}|$ and a system with diversity order $qM_r$, then the
rate $R$ that can be achieved is given by
\begin{equation}
\label{eq:RateBndTSC}
R \leq (M_t - q + 1)
\end{equation}
in symbols per transmission, {\em i.e.,} the rate is
$R\log_2|\mathcal{A}|$ bits per transmission.
\end{theorem}
Just as Theorem \ref{thm:RateDivTSC} shows the trade-off between
achieving high-rate and high-diversity given a fixed transmit alphabet
constraint for a flat fading channel, there also exists a trade-off
between achievable rate and diversity for frequency selective
channels, and we aim to characterize this trade-off\footnote{It is
tempting to guess that the trade-off for the fading ISI case is just a
$(\nu+1)-$fold increase in the diversity order.}. A corollary will be
an upper bound on the performance of diversity embedded codes for ISI
channels.  This can be seen by observing that we can easily extend the
proof in \cite{TSC98,LuKumar03} to the case where we have the Toeplitz
structure as given in (\ref{eqn:model1}). Note that the diversity
order of the codes for fading ISI channel is given by the rank of the
corresponding (Toeplitz) codeword difference matrix. Since this rank
is upper bounded as seen in (\ref{upper_bound}), we see that we
immediately obtain a trivial upper bound for the rate-diversity
trade-off for th fading ISI case as follows.

\begin{lemma}
\label{lem:RateDivISIUppBnd}
If we use a constellation of size $|\mathcal{A}|$ and the diversity
order of the system is $q(\nu+1)M_r$, then the rate $R$ in symbols per
transmission that can be achieved is upper bounded as
\begin{equation}
\label{eq:RateBndISIUppBnd}
R \leq (M_t - q + 1).
\end{equation}
\end{lemma}
Note that in Theorem \ref{thm:RDisi}, we establish a corresponding
lower bound that asymptotically (in block size) matches this upper
bound. Note that due to the zero padding structure for ISI channels,
the effective rate $R^{eff}$ is going to be smaller than the rate of
space-time code. Since we do not utilize $\nu$ transmissions over a
block of $T$ transmissions for each of the antennas we can only hope
for a rate $R$ asymptotically in transmission block size $T$.

\subsection{Set Partitioning of QAM and QPSK Constellations}
\label{subsec:setpart}
Let $\Gamma_1,\ldots,\Gamma_L$ be a $L$-level
partition where $\Gamma_i$ is a refinement of partition
$\Gamma_{i-1}$. We view this as a rooted tree, where the root is the
entire signal constellation and the vertices at level $i$ are the
subsets that constitute the partition $\Gamma_i$. In this paper we
consider only binary partitions, and therefore subsets of partition
$\Gamma_i$ can be labeled by binary strings $a_1,\ldots,a_i$, which
specify the path from the root to the specified vertex.

\begin{figure}
\begin{center}
\includegraphics[scale=0.35]{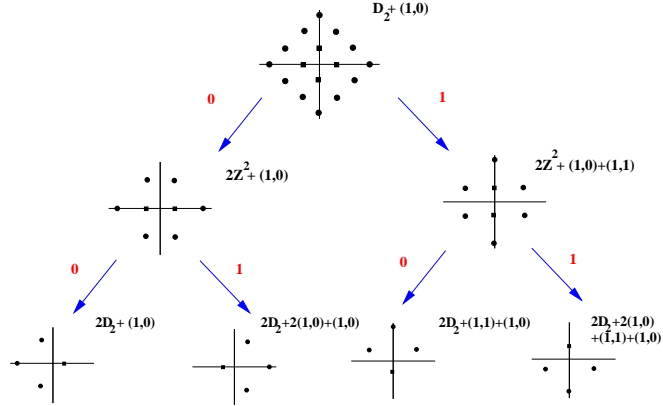}
\end{center}
\caption{A binary partition of a QAM constellation}
\label{fig:binpart}
\end{figure}

Signal points in QAM constellations are drawn from some realization of the
integer lattice $\ZZ^2$.
We focus on the particular
realization shown in Figure \ref{fig:binpart}, where the integer lattice has
been scaled by {\small$\left [ \begin{array}{cc} 1 &1 \\ 1 &-1
\end{array} \right ]$}
to give the lattice $D_2=\{(a,b)|a,b\in \ZZ , a+b\equiv 0
(\mathrm{mod} 2)\}$, and then translated by $(1,0)$. The constellation
is formed by taking all the points from $\Lambda$ that fall within a
bounding region $\mathcal{R}$. The size of the constellation is
proportional to the area of the bounding region, and in Figure
\ref{fig:binpart}, the bounding region encloses $16$ points.

Binary partitions of QAM constellations are typically based on the following chain
of lattices
\begin{displaymath}
D_2 \supset 2\ZZ^2 \supset 2D_2 \supset 4\ZZ^2 \supset \ldots 2^{i-1}D_2\supset
2^i\ZZ^2 \supset 2^iD_2 \supset \ldots
\end{displaymath}

In Figure \ref{fig:binpart}, the subsets at level 1 are, to within
translation, cosets of $2\ZZ^2$ in $D_2$ and the subsets at level 2
are cosets of $2D_2$.  In general the subsets at level $2i$ are pairs
of cosets of $2^iD_2$ where the union is a coset of $2^i\ZZ^2$, and
the subsets at level $2i+1$ are pairs of cosets of $2^{i+1}\ZZ^2$
where the union is a coset of $2^iD_2$. Note that implicit in Figure
\ref{fig:binpart} is a binary partition of QPSK, where the points
$1,-1,i,-i$ are labeled $00,01,11,10$ respectively.  Binary
partitions of PSK constellations are described in Section
\ref{subsec:AlgProp}.

\hspace{0in}

\subsection{Algebraic properties of binary partitions}
\label{subsec:AlgProp}

\begin{figure}[h]
\begin{center}
\includegraphics[scale=0.6]{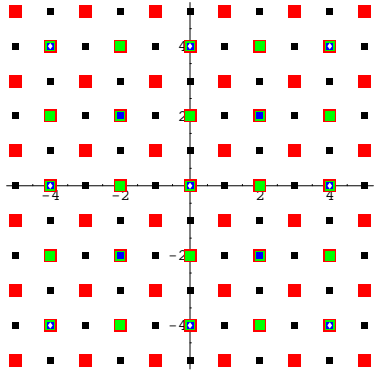}
\caption{Decomposition for QAM constellations}
\label{fig:qam}
\end{center}
\end{figure}

The QAM constellations can be represented through a lattice chain
$\Lambda|\Lambda_1|\Lambda_2|\ldots$, where $\Lambda=\ZZ^2$ is the
integer lattice. The lattices in the chain are produced with the
generator matrix $\Gbf^k$ where $\Gbf = \left [\begin{array}{cc} 1 &1
\\ 1 &-1 \end{array}\right ]$. Given this, we can represent the
$2^k$-QAM constellation as $\Lambda|\Lambda_k$, {\em i.e.,} the coset
representatives of $\Lambda$ in $\Lambda_k$.  The lattice $\Lambda$
can also be written as the set of Gaussian integers $Z[i]=\{a + b i:
a,b\in\ZZ\}$. Similarly we can write the lattice $\Lambda_k$ as
$\{(a+b i)(1-i)^k: a,b\in\ZZ\}$.  This decomposition of the QAM
constellation is illustrated in Figure \ref{fig:qam}.  Therefore,
using this we can represent any point $s$ in a $2^L$-QAM constellation
using a $L$-length bit string as
\begin{equation}
\label{eq:QAMrep1}
s - c(L)\equiv\sum_{l=0}^{L-1} b_l (1-i)^l \hspace{0.2in} \mbox{mod } (1-i)^{L},
\end{equation}
where we define $f \equiv g \hspace{0.2in}\mbox{mod } (1-i)^{L}$ if there exist
$c,d\in \ZZ$ such that $f = (c+d i)(1-i)^{L} + g$. Also in (\ref{eq:QAMrep1})
the constant $c(L)=\frac{1}{2}$ for odd $L$ and $\frac{1}{2}(1+i)$ for even $L$.

Binary partitions of PSK constellations are based on a chain of
subfields of the cyclotomic field $\mathbb{Q}(\xi_{2^L})$ obtained by
adjoining $\xi_{2^L}=\exp(2\pi i/2^L)$ to the rational field
$\mathbb{Q}$.
Analogous to (\ref{eq:QAMrep1}), points in the $2^L$-PSK
constellation can be represented as
\begin{equation}
\label{eq:PSKrep}
s = \prod_{l=0}^{L-1}
(\xi^{2^l})^{b_l},
\end{equation}
where $\xi=\xi_{2^L}=\exp(2\pi i/2^L)$ and $\xi^{2^L}$ is a primitive
element for $\mathbb{Q}(\xi_{2^{L-l}})$. Note that $1-\xi$ is prime
in $\mathbb{Z}[\xi]$ and the quotient $\mathbb{Z}[\xi]/(1-\xi)$ is the
field $\mathbb{Z}_2$.

The field $\mathbb{Q}(\xi_{2^L})$ is a degree $2^{L-1}$ extension of
$\mathbb{Q}$.  Every rational number is a quotient $a/b$, where
$a,b\in\ZZ$, and every complex number in $\mathbb{Q}(i)$ is a quotient
$a/b$, where $a,b$ are Gaussian integers.  In general every complex
number in $\mathbb{Q}(\xi_{2^L})$ is a quotient $a/b$, where $a,b$ are
integer linear combinations of
$1,\xi_{2^L},\ldots,\xi_{2^L}^{2^{L-1}-1}$ and $b\neq 0$. For more
details about cyclotomic fields see \cite{Washington}.  Note that
$\xi_{2^L}^{2^{L-1}}=-1$, so that
$\xi_{2^L}^j=-\xi_{2^L}^{2^{L-1}+j}$, for $j=0,1,\ldots,2^{L-1}-1$.

We have a chain of fields
\begin{displaymath}
\mathbb{Q} = \mathbb{Q}(\xi_{2}) \subset \mathbb{Q}(i) = \mathbb{Q}(\xi_{4})
\subset \mathbb{Q}(\xi_{8}) \ldots \subset \mathbb{Q}(\xi_{2^L}).
\end{displaymath}
These observations can be used to establish the performance of the
multi-level diversity embedded codes \cite{DDCA05,DCDA06}.

\section{Diversity embedded codes for ISI channels}
\label{sec:MLcons}
In this section we will first recall the construction of multi-level
(non-linear) space-time codes for transmission over {\em flat fading}
channels that are matched to a binary partition of a QAM or PSK
constellation (see \cite{DCDA06}). We will give the construction and
refer the reader to \cite{DCDA06} for proofs of code performance for
the flat-fading case. Following this we will use the structure imposed
by the ISI on the space time code as in (\ref{eqn:model1}) to
construct multilevel codes {\em suitable for ISI channels} using
binary matrices which are constructed in Section \ref{sec:Rank}.

\subsection{Multi-Level Constructions for Flat Fading Channels}
\label{subsec:Code_Flat}
Given an L-level binary partition of a QAM or PSK signal
constellation, a space-time codeword is an array $\Kbf = \{\Kbf_1,
\Kbf_2,\ldots, \Kbf_L\}$ determined by a sequence of binary matrices,
where matrix, $\Kbf_i$ specifies the space-time array at level $i$. A
{\it multi-level space-time code} is defined by the choice of the
constituent sets of binary matrices $\mathcal{K}_1,
\mathcal{K}_2,\ldots, \mathcal{K}_L$. These sets of binary matrices
provide rank guarantees necessary to achieve the diversity orders
required for each message set. For $i=1, \ldots, L$ the binary matrix
$\Kbf_i$ is required to be in the set $\mathcal{K}_i$.

\begin{figure}[h]
\centering \input{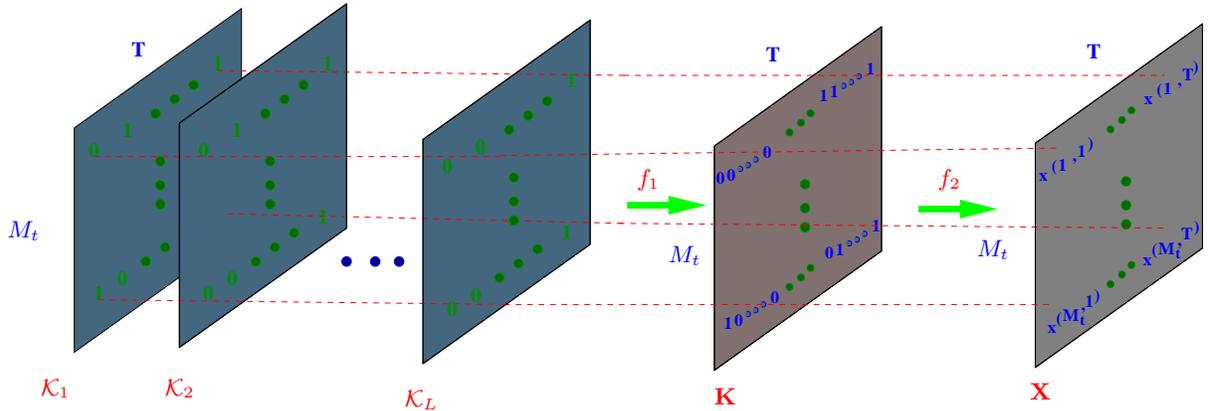}
\caption{Schematic representation of the multi-level code construction.}
\label{fig:CodeConst}
\end{figure}

Given message sets $\{\mathcal{E}_i\}_{i=1}^L$, they are
mapped to the space-time codeword $\Xbf$ as shown below.
\begin{eqnarray}
\label{eq:FFdivembMLC}
\hspace{-0.1in} \{\mathcal{E}\}_{i=1}^L
\stackrel{f_1}{\longrightarrow} \Kbf = \left [ \begin{array}{ccc}
K(1,1) &\ldots &K(1,T)\\ \vdots &\vdots &\vdots \\ K(M_t,1) &\ldots
&K(M_t,T) \end{array} \right ] \stackrel{f_2}{\longrightarrow} \Xbf =
\left [ \begin{array}{ccc} x(1,1) &\ldots &x(1,T)\\ \vdots &\vdots
&\vdots \\ x(M_t,1) &\ldots &x(M_t,T)\end{array} \right ]
\end{eqnarray}
where the matrix $\Kbf$ is specified by $K(m,n) \in
\{0,1\}^{\log(|\mathcal{A}|)}$ {\em i.e.,} binary string and
$x(m,n)\in \mathcal{A}$. This construction is illustrated in Figure
\ref{fig:CodeConst} for a constellation size of $L$ bits.

In summary, given the message set, we first choose the matrices
$\Kbf_1,\ldots.\Kbf_L$.  The first mapping $f_1$ is obtained by taking
matrices and constructing the matrix $\Kbf\in\Complex^{M_t\times T}$
each of whose entries is constructed by concatenating the bits from
the corresponding entries in the matrices $\Kbf_1,\ldots.\Kbf_L$ into
$L$-length bit-string. This matrix is then mapped to the space-time
codeword through a constellation mapper $f_2$, for example the
set-partition mapping given in Section \ref{subsec:setpart}. Using
this sequence of $L$ matrices, we obtain the space-time codeword as
seen in Figure
\ref{fig:CodeConst}.

For flat fading channels the sets $\mathcal{K}_l,l=1,\ldots,L$ are
binary $M_t\times T$ matrices such that for any distinct pair of
matrices $\Abf, \Bbf \in \mathcal{K}$ the rank of $\Abf-\Bbf$ is at
least $M_t-d$. The size of $\mathcal{K}$ is at most $2^{(d+1)T}$
since the first $d+1$ rows of $\Abf$ and $\Bbf$ must be distinct, and
there is a classical example \cite{Gabidulin} that achieves the bound
(this construction was also given in \cite{LuKumar03,LuKumar05}).

With the rate achieved on the $l^{th}$ layer defined as
$R_l=\frac{1}{T}\log|\mathcal{K}_{l}|$ it can be shown \cite{DCDA06}
that this construction for QAM constellations achieves the
rate-diversity tuple $(R_1,M_r d_1(\nu+1),\ldots,$ $R_L,M_r
d_L(\nu+1))$, with the overall equivalent single layer code achieving
rate-diversity point, $(\sum_l R_l,M_rd_L(\nu+1))$. Optimal decoding
employs a maximum-likelihood decoder which jointly decodes the message
sets. This is the decoder for which the performance is summarized in
Theorem \ref{thm:MLC_Flat}.
\begin{theorem}\cite{DCDA06}
\label{thm:MLC_Flat}
Let $\mathcal{C}$ be a multi-level space-time code for a QAM or
$M$-PSK constellation of size $2^L$ with $M_t$ transmit antennas that
is determined by constituent sets of binary matrices
$\mathcal{K}_l,l=1,\ldots,L$ with binary rank guarantees $d_1 \geq d_2
\ldots \geq d_L$. For joint maximum-likelihood decoding, the input
bits that select the codeword from the $i$th matrix $\mathcal{K}_i$
are guaranteed diversity $d_i M_r $ in the complex domain when
transmitted over a flat fading channel.
\end{theorem}

\subsection{Multi-Level Construction for ISI Channels}
\label{subsec:Code_ISI}
In this section we use the idea of multi-level diversity embedded
codes for flat fading channels as in Section \ref{subsec:Code_Flat}
and the structure imposed by the ISI on the space time code as in
(\ref{eqn:model1}) to motivate construction and analysis of a class of
binary matrices as follows.

We apply the idea suggested by the constructions of multi-level codes
for flat-fading channels to the fading ISI case. We do this by
applying a zero padding as seen in (\ref{eq:X1def}) along with
mappings of binary matrices to the transmit signal alphabet. That is,
we use the mapping given in (\ref{eq:FFdivembMLC}) for a block size of
$T$ with the constraint that the last $\nu$ entries of the mapping
lead to {\em given} alphabets (taken to be zero without loss of
generality). This is combined with binary sets $\mathcal{K}_{\nu,d}$,
which we specify in definition \ref{suitable_isi}. This means that
over a time period $T$, we transmit a sequence ${\bf x}[0], {\bf
x}[1], \ldots, {\bf x}[T-\nu-1]$ which are mapped from the inputs bits
using a structure given in (\ref{eq:FFdivembMLC}). Therefore, given that
we transmit the sequence shown in (\ref{eq:XisiDef}),
\begin{align}
\label{eq:XisiDef}
\mathbf{X}^{(1)} &=  \left[  
\begin{array}{ccccccc}
{\bf x}[0] & {\bf x}[1] & \ldots  & {\bf x}[T-\nu-1] & 0    & \ldots         & 0 
\end{array}
\right ],
\end{align}
we need a mapping from a binary matrix as in
(\ref{eq:FFdivembMLC}). For a constellation of size $2^L$, we do this
by taking message sets $\{\mathcal{E}_i\}_{i=1}^L$ and mapping them to
a codeword with the structure given in (\ref{eq:XisiDef}) as follows,
\begin{small}
\begin{eqnarray}
\label{eq:ISIdivembMLC}
\hspace{-0.1in}
\{\mathcal{E}_i\}_{i=1}^L \stackrel{f_1}{\longrightarrow} \Kbf^{(1)} =  \left [ \begin{array}{ccc}
K(1,1) &\ldots &K(1,T)\\ \vdots &\vdots &\vdots \\
K(M_t,1) &\ldots &K(M_t,T) \end{array} \right ] \stackrel{f_2}{\longrightarrow}
\Xbf^{(1)} = \left[ 
\begin{array}{ccccccc}
{\bf x}[0] & {\bf x}[1] & \ldots  & {\bf x}[T-\nu-1] & 0    & \ldots         & 0 
\end{array}
\right ]
\end{eqnarray}
\end{small}
where the $(m,n)^{th}$ entry of $\Kbf^{(1)}$ is given by $K(m,n) \in
\{0,1\}^{\log(|\mathcal{A}|)}$ {\em i.e.,} binary string. Since the mapping $f_2$
is just the set-partitioning mapping specified in Section \ref{subsec:setpart},
we need the last $\nu$ columns of $\Kbf^{(1)}$ to be {\em given constants} for {\em all}
choices of the message sets $\{\mathcal{E}_i\}_{i=1}^L$. That is, we need the following
structure for the matrix $\Kbf^{(1)}$, 
\begin{align}
\mathbf{K}^{(1)} &=  \left[  
\begin{array}{ccccccc}
{\bf k}[0] & {\bf k}[1] & \ldots & {\bf k}[T-\nu-1] & \mathbf{0} &
\ldots & \mathbf{0}
\end{array}
\right], 
\label{eq:KisiDef}
\end{align}
where, as before, $\{{\bf k}[i]\}$ are columns of binary strings of length
$L$, and with no loss of generality, we have specified the last $\nu$
columns of $\Kbf^{(1)}$ to be the zero strings. 

Given that we have an ISI channel, the transmitted codeword with the
structure given in (\ref{eq:XisiDef}) gives an equivalent codeword
matrix with a Toeplitz structure, as specified in (\ref{eqn:model1}). This Toeplitz
structure is equivalent to mapping a Toeplitz matrix $\Kbf$ of binary strings with
the structure
\begin{align}
\label{eqn:ToepKmap}
\Kbf & = 
\left[
\begin{array}{ccccccc}
{\bf k}[0] & {\bf k}[1] & \ldots & {\bf k}[T-\nu-1] & \mathbf{0} &
\ldots & \mathbf{0} \\ \mathbf{0} & {\bf k}[0] & {\bf k}[1] & \ldots &
{\bf k}[T-\nu-1] & \mathbf{0} & \mathbf{0} \\ \ldots & & \vdots &
\ddots & .  & .  & \vdots\\ \mathbf{0} & \ldots & \mathbf{0} &{\bf
k}[\mathbf{0}] & {\bf k}[1] & \ldots & {\bf k}[T-\nu-1]
\end{array}
\right],
\end{align}
to $\Xbf$ using the constellation mapping $f_2$. Therefore, as in the
flat fading case, given the message set, we first choose the binary
matrices $\Kbf_1^{(1)},\ldots.\Kbf_L^{(1)}$, each of which have the
structure specified below in (\ref{eq:Bdef}). These put together give
us the matrix of binary strings $\Kbf^{(1)}$. This in turn, due to the
ISI channel, is relates to $\Kbf$, the Toeplitz matrix of binary
strings, given above in (\ref{eqn:ToepKmap}). Therefore, the choice of
matrices $\Kbf_1^{(1)},\ldots.\Kbf_L^{(1)}$, for the ISI case,
naturally is equivalent to a choice of Toeplitz binary matrices,
$\Kbf_1,\ldots.\Kbf_L$, as specified in (\ref{eqn:mapping}) below.

Therefore, for the multi-level coding structure we have used,
analogous to the flat fading case studied in \cite{DCDA06}, we need to
study the rank properties of sets of binary Toeplitz matrices as
specified below.  Consider the matrix $\mathbf{B}\in\base^{M_t\times
T}$, with the following structure,
\begin{align}
\mathbf{B} &=  \left[  
\begin{array}{ccccccc}
{\bf c}[0] & {\bf c}[1] & \ldots & {\bf c}[T-\nu-1] & \mathbf{0} &
\ldots & \mathbf{0}
\end{array}
\right], 
\label{eq:Bdef}
\end{align}
where $\mathbf{c}[n]\in\mathbb{F}_2^{M_t\times
1},\,\,n=0,\ldots,T-\nu-1$.  We define a mapping
$\Theta:\base^{M_t\times T}\rightarrow \base^{(\nu+1)M_t\times T} $
by,
\begin{align}
\label{eqn:mapping}
\Theta\left(\mathbf{B}\right) & = 
\left[
\begin{array}{ccccccc}
{\bf c}[0] & {\bf c}[1] & \ldots & {\bf c}[T-\nu-1] & \mathbf{0} &
\ldots & \mathbf{0} \\ \mathbf{0} & {\bf c}[0] & {\bf c}[1] & \ldots &
{\bf c}[T-\nu-1] & \mathbf{0} & \mathbf{0} \\ \ldots & & \vdots &
\ddots & .  & .  & \vdots\\ \mathbf{0} & \ldots & \mathbf{0} &{\bf
c}[\mathbf{0}] & {\bf c}[1] & \ldots & {\bf c}[T-\nu-1]
\end{array}
\right]
\end{align}

\begin{definition}
\label{suitable_isi}
Define
$\mathcal{K}_{\nu,d}\subset\{\mathbf{B}:\mathbf{B}\in\base^{M_t\times
T}\}$ to be the set of binary matrices of the form given in
(\ref{eq:Bdef}) if for some fixed $T_{thr}$ they satisfy the following
properties for $T\geq T_{thr}$.
\begin{itemize}
\item For any distinct pair of matrices $\Abf, \Bbf \in \mathcal{K}_{\nu,d}$
the rank of $\left[\Theta(\Abf)-\Theta(\Bbf)\right ]$ is at least $d(\nu+1)$.
\item $|\mathcal{K}_{\nu,d}|\geq 2^{T(M_t-d+1)-\nu M_t}$.
\end{itemize}
\end{definition}
Note that in Section \ref{subsec:Code_Flat} the first step in code
construction was constructing the sets $\mathcal{K}_l,l=1,\ldots,L$
from which the matrices $\Kbf_1,\ldots.\Kbf_L$ were chosen. In the
case of flat fading channels there are constructions by
\cite{Gabidulin} but these do not satisfy the rank guarantee
properties in Definition \ref{suitable_isi}. We will postpone the
construction of such sets of binary matrices $\{\mathcal{K}_{\nu,d}\}$
to Section \ref{sec:Rank}, where we show that we can set $T_{thr}= R
\nu +(M_t-1)(\nu+1) (2^{R}-1)$. More formally, in Section  \ref{sec:Rank}, we show
that,

\begin{lemma}
\label{lem:KsetCons}
For block size $T\geq T_{thr}= R \nu +(M_t-1)(\nu+1) (2^{R}-1)$, there
exist sets of binary matrices $\mathcal{K}_{\nu,d}$ which satisfy the
requirements of Definition \ref{suitable_isi}.
\end{lemma}

Adapted easily from \cite{DCDA06} we can state the formal construction
guarantee for the diversity embedded code for transmission over the
ISI channel as follows.
\begin{theorem}
\label{thm:MLC_ISI}
Let $\mathcal{C}$ be a multi-level space-time code for a QAM or PSK
constellation of size $2^L$ with $M_t$ transmit antennas that is
determined by constituent sets of binary matrices
$\mathcal{K}_l=\mathcal{K}_{\nu,d_l},l=1,\ldots,L$, such that $d_1\geq
d_2 \ldots \geq d_L$. For joint maximum-likelihood decoding, the input
bits that select the codeword from the $l$th set $\mathcal{K}_l$ are
guaranteed diversity $d_l(\nu+1)M_r $ in the complex domain when
transmitted over an ISI channel with $\nu+1$ taps.
\end{theorem}
The proof of the Theorem \ref{thm:MLC_ISI} follows from the same
techniques as in \cite{DCDA06} by mapping binary matrices with desired
rank guarantees to rank guarantees in complex domain. In particular,
given sets of (Toeplitz) binary matrices
$\mathcal{K}_l=\mathcal{K}_{\nu,d_l},l=1,\ldots,L$, which have rank
guarantees of $\{d_l\}$, given the set-partitioning mapping $f_2$, we
can lift the binary rank properties to the complex domain. Therefore,
the main challenge, addressed in this paper, is the construction of
such sets of binary matrices with rank guarantees.

Therefore the codewords from $l$th layer achieve a rate
$R_l=\frac{1}{T}\log|\mathcal{K}_{\nu,d_l}|$ and diversity order
$d_l(\nu+1)M_r $.  From Definition \ref{suitable_isi} it follows that
the size of $\mathcal{K}_{\nu,d_l}$ can be made at least as large as
$2^{T\left(M_t - d_l + 1 \right)- \nu M_t}$. Similar to \cite{DCDA06}
this construction for QAM constellations achieves the rate-diversity
tuple $(R_1,M_rd_1(\nu+1),\ldots,R_L,M_rd_L(\nu+1))$, with the overall
equivalent single layer code achieving rate-diversity point, $(\sum_l
R_l,M_rd_L(\nu+1))$.

In particular, we can construct a space-time code by choosing
identical diversity requirements for all the layers, {\em i.e.},
$d_1=d_2=\ldots=d_L$. From this we conclude that the rate diversity
tradeoff for the ISI channel can be characterized as follows:

\begin{theorem}{\bf (Rate Diversity Tradeoff for ISI Channels)}
\label{thm:RDisi}
Consider transmission over a $\nu$ tap ISI channel with $M_t$ transmit
antennas from a QAM or PSK signal constellation $\mathcal{A}$ with
$|\mathcal{A}|=2^L$ and communication over a time period $T$ such that
$ T \geq T_{thr}$. For diversity order $d_{isi}=d(\nu+1)M_r$, the rate
diversity tradeoff is given by,
\begin{align*}
(M_t-d+1) -\frac{\nu }{T}M_t & \leq R^{eff} \leq \left( M_t- d +1 \right).
\end{align*}
\end{theorem}
The lower bound follows directly from Theorem \ref{thm:MLC_ISI} and
the upper bound follows from lemma \ref{lem:RateDivISIUppBnd}.  Note
that the bounds in the above theorem are tight as $T\rightarrow
\infty$.

\section{Diversity Embedded Trellis Codes}
\label{sec:Trellis}
The construction of diversity embedded trellis codes for ISI channels
is quite similar to the construction of block codes. Again the idea is
to construct binary convolutional codes with the following properties.

\begin{definition}
\label{suitable_isi_trellis}
Define
$\mathcal{P}_{\nu,d}\subset\{\mathbf{B}:\mathbf{B}\in\base^{M_t\times
T}\}$ to be the set of binary matrices of the form given in
(\ref{eq:Bdef}) if for some fixed $T_{thr}$ they satisfy the following
properties for $T\geq T_{thr}$
\begin{itemize}
\item For any distinct pair of matrices $\Abf, \Bbf \in \mathcal{P}_{\nu,d}$
the rank of $\Gamma(\Abf)-\Gamma(\Bbf)$ is at least $d(\nu+1)$.
\item $\log|\mathcal{P}_{\nu,d}|\geq R(
T-\nu-(\nu+1)(M_t-1)(2^R-1)(2^{R-1}))$, where $R=M_t-d+1$.
\end{itemize}
\end{definition}
Using these sets of matrices obtained by appropriately choosing the
underlying convolutional codes the diversity embedding properties are
ensured.

We will postpone the construction of such sets of binary matrices to
Section \ref{subsec:ConvCode} where using Lemma \ref{lem:KsetCons} along
with particular choices of convolutional codes we show the following result.

\begin{lemma}
\label{lem:PsetCons}
For block size $T\geq T_{thr}= (2^R-1)\nu + (2^R-1)(\nu+1) \left(
(M_t-2) (2^{R}-1)+R\right )$, where $R=M_t-d+1$, there exist sets of binary
matrices $\mathcal{P}_{\nu,d}$ which satisfy the requirements of
Definition \ref{suitable_isi_trellis}.
\end{lemma}

As in the case of block codes in Section \ref{sec:MLcons}, given a
$L$-level binary partition of a QAM or PSK signal constellation, a
diversity embedded convolutional space-time codeword is defined by an
array $\Pbf = \{\Pbf^{(1)}, \Pbf^{(2)},\ldots, \Pbf^{(L)}\}$
determined by a sequence of binary matrices, where matrix,
$\Pbf^{(i)}$ specifies the space-time array at level $i$. Adapted
easily from \cite{DCDA06} we can state the formal construction
guarantee for the diversity embedded trellis code for ISI channels as
follows.

\begin{theorem}
\label{thm:MLC_trellis}
Let $\mathcal{C}$ be a multi-level space-time code for a QAM or PSK
constellation of size $2^L$ with $M_t$ transmit antennas that is
determined by constituent sets of binary matrices
$\mathcal{P}_l=\mathcal{P}_{\nu,d_l},l=1,\ldots,L$, such that $d_1\geq
d_2 \ldots \geq d_L$. For joint maximum-likelihood decoding, the input
bits that select the codeword from the $l$th set
$\mathcal{P}_l$ are guaranteed diversity $d_l(\nu+1)M_r $ in the complex
domain. 
\end{theorem}

The proof of the Theorem \ref{thm:MLC_trellis} follows from the same
techniques as in \cite{DCDA06} by mapping binary matrices with desired
rank guarantees to rank guarantees in complex domain.  As in the proof
of Theorem \ref{thm:MLC_ISI}, the main difficulty is in constructing
these sets of binary matrices with thee given rank guarantees, using
convolutional codes.  We give such a construction in Section
\ref{subsec:ConvCode}.  Therefore the codewords from $l$th layer
achieve a rate $R_l=\frac{1}{T}\log|\mathcal{P}_{\nu,d_l}|$ and
diversity order $d_l(\nu+1)M_r $.  From Definition
\ref{suitable_isi_trellis} it follows that the size of
$\mathcal{P}_{\nu,d_l}$ can be made at least as large as
$2^{R(T-\nu-(\nu+1)(M_t-1)(2^R-1)(2^{R-1}))}$, which in the limit as
$T\rightarrow \infty$ tends to $2^R$.  Similar to \cite{DCDA06} this
construction for QAM constellations achieves the rate-diversity tuple
$(R_1,M_rd_1(\nu+1),\ldots,R_L,M_rd_L(\nu+1))$, with the overall
equivalent single layer code achieving rate-diversity point, $(\sum_l
R_l,M_rd_L(\nu+1))$.

\begin{figure}[t]
\input{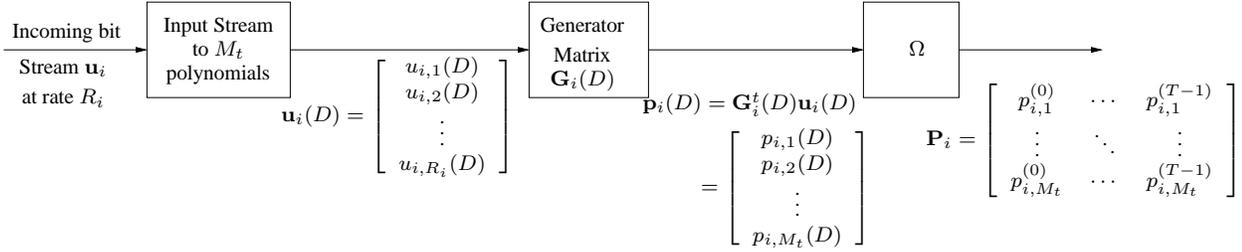}
\vspace{1cm}
\caption{Binary matrices for each layer}
\label{fig:DivEmbTrel}
\end{figure}

We illustrate the idea by examining the construction for each of the
layers. The construction is shown in Figure
\ref{fig:DivEmbTrel}. Given the input stream for each layer $i$, the
first block in the figure maps these inputs to the coefficients of
$R_i$ polynomials $u_{i,j}(D), j=1,\ldots,R_i$ in $\base [D]$. The
second block multiplies the input vector $\{u_{i,j}(D)\}_{j=1}^{R_i}$
by the generator matrix ${\bf G}_i(D)$, with special structure which
we define in the following subsection, and generates a vector ${\bf
u}_i(D)$ of polynomials. The final block $\Omega$ then maps this
vector ${\bf p}_i(D)$ to a binary matrix ${\bf P}_i\in \base^{M_t\times
T}$.

We define the set $\mathcal{P}_i=\mathcal{P}_{\nu,i}$ to be the set of
all output matrices ${\bf P}_i$ for all possible inputs on the stream
$i$. Note that these sets satisfy the properties in Definition
\ref{suitable_isi_trellis}.

\subsection{Binary Convolutional Codes}
\label{subsec:ConvCode}
Explicit construction of full diversity maximum rate binary
convolutional codes was first shown in \cite{HE00}. This was extended
for general points on the rate diversity tradeoff for flat fading
channels in \cite{LuKumar05}. We will give constructions for such sets
of binary matrices for ISI channels in this section.

Consider the construction for a particular layer above. We will see
the construction of rate $R$ symbols per transmission, and rank
distance of $(\nu+1)(M_t-R+1)$ binary codes for transmission over the
ISI channel. Represent the generator matrix or transfer function
matrix ${\bf G}$ for this code by an $R\times M_t$ generator matrix
given by,
\begin{align}
{\bf G} & = \left[ \begin{array}{cccc}
g_1^{(1)}(D) & g_1^{(2)}(D) & \cdots & g_1^{(M_t)}(D) \\
g_2^{(1)}(D) & g_2^{(2)}(D) & \cdots & g_2^{(M_t)}(D) \\
\vdots & \ddots & \ddots & \vdots \\
g_{R}^{(1)}(D) & g_{R}^{(2)}(D) & \cdots & g_{R}^{(M_t)}(D)
\end{array}
\right] 
\end{align}
Denoting $\xi=D^{(\nu+1)(2^R-1)}$ we choose 
\begin{equation}
\label{eq:PolyChoice}
g_{l}^{(q)}(D)=\xi^{(q-1) 2^{(l-1)}}.
\end{equation}

The input message polynomial is represented by the vector
of message polynomial
\begin{align}
{\bf u}(D) & =  \left[ \begin{array}{cccc} u_{1}(D) & u_{2}(D) &
\cdots  & u_{R}(D)
\end{array}
\right]^t
\end{align}
where $u_{i}(D) \in \base[D] $. The code polynomial vector is given by
\begin{align}
\label{eqn21}
{\bf p}(D) & =  {\bf G}^t (D) {\bf u}(D) \nonumber \\
& =  \left[ \begin{array}{cccc} p_{1}(D) & p_{2}(D) & \cdots & p_{M_t}(D) 
\end{array}
\right]^t. 
\end{align}

The $M_t \times T $ code matrix which is actually transmitted on the
antenna is given by
\begin{align}
{\bf P} & =  \left[ \begin{array}{ccc}
p_1^0 & \cdots & p_1^{T-1} \\
\vdots & \ddots & \vdots \\
p_{M_t}^0 & \cdots & p_{M_t}^{T-1}
\end{array}
\right]
\end{align}
where $p_{i}^j$ is the $j^{th}$ coefficient of the polynomial $p_{i}$
in (\ref{eqn21}). We make a distinction between ${\bf p}(D)$ which is
a vector of polynomials in $D$ and ${\bf P}$ which is a binary
matrix. This mapping is denoted by $\Omega$ {\em i.e.} $\Omega({\bf
p}(D))={\bf P}$

Note that in order that the matrix ${\bf P}$ satisfies the structure
in (\ref{eq:Bdef}) we require the $\nu$ largest coefficients of each
$p_{i}(D)$ in (\ref{eqn21}) to be zero, {\em i.e.},
\begin{align}
p_{i}^j & =0 \qquad \forall \, i\in\{1,\ldots,M_t\} \ {\rm and}\ \forall \, j \in \{T-\nu-1,\ldots,T-1\}.
\label{eqn:zero_padding}
\end{align}
With this constraint we get that, 
\begin{align*}
deg(u_i(D)) & \leq T-1-\nu - \max_{u,v} deg(g_u^{(v)}) \\ 
& = T - 1 -\nu- (\nu+1)(M_t-1)(2^R-1)(2^{R-1})
\end{align*}
where the last equality follows from our particular choice of
$g_u^{(v)}$ given in (\ref{eq:PolyChoice}). Note that this
convolutional code corresponds to a effective rate of
\begin{align}
R^{eff} & = \frac{\log \left( 2^{T-1-\nu-(\nu+1)(M_t-1)(2^R-1)(2^{R-1}) +1 }\cdot 2^R \right) }{T} \nonumber \\
& =  \frac{R\left( T-\nu-(\nu+1)(M_t-1)(2^R-1)(2^{R-1}) \right)}{T}\quad bits/Tx
\end{align}
which asymptotically tends to $R$ as $T\rightarrow \infty $.

Also, observe that, 
\begin{align}
\Theta( {\bf P}) & = \left[ 
\begin{array}{c}
\Omega({\bf p}(D))\\
\Omega(D {\bf p}(D))\\
\vdots \\
\Omega( D^{\nu} {\bf p}(D))
\end{array}
\right].
\end{align}
From this we can conclude that,
\begin{align*}
\Theta( {\bf P})) & = \Omega \left( \tilde{{\bf G}}^t (D) {\bf u}(D) \right)
\end{align*}
where $\tilde{{\bf G}} \in \base^{R \times (\nu+1)M_t }$ is given by,
{\footnotesize
\begin{align}
\tilde{{\bf G}} & = \left[ \begin{array}{ccccccccccc} g_1^{(1)}(D) &
g_1^{(2)}(D) & \cdots & g_1^{(M_t)}(D) & D g_1^{(1)}(D) & \cdots & D
g_1^{(M_t)}(D) & \cdots & D^{\nu} g_1^{(1)}(D) & \cdots & D^{\nu}
g_1^{(M_t)}(D) \\ g_2^{(1)}(D) & g_2^{(2)}(D) & \cdots &
g_2^{(M_t)}(D) & D g_2^{(1)}(D) & \cdots & D g_2^{(M_t)}(D) & \cdots &
D^{\nu} g_2^{(1)}(D) & \cdots & D^{\nu} g_2^{(M_t)}(D)\\ \vdots &
\ddots & \ddots & \vdots & & & & & & & \\ g_{R}^{(1)}(D) &
g_{R}^{(2)}(D) & \cdots & g_{R}^{(M_t)}(D) & D g_R^{(1)}(D) & \cdots &
D g_R^{(M_t)}(D) & \cdots & D^{\nu} g_R^{(1)}(D) & \cdots & D^{\nu}
g_R^{(M_t)}(D)
\end{array}
\right] 
\end{align}
} With our particular choice of $g_l^{(q)}(D)$, given in
(\ref{eq:PolyChoice}), we can write this as,
\begin{align}
\label{eqn:Grepr_trellis}
\tilde{{\bf G}}^t & = 
\left[
\begin{array}{cccc}
1 & \ldots & 1 & 1 \\
\xi & \xi^2 & \ldots & \xi^{2^{R-1}} \\
\xi^{2} & (\xi^{2})^2 &  \ldots & (\xi^{ 2 })^{2^{R-1} } \\
\vdots & & \vdots & \\
\xi^{(M_t-1)} & (\xi^{(M_t-1)})^2 & \ldots & (\xi^{(M_t-1)} )^{2^{R-1} }  \\
D & \ldots & D & D \\
 & \vdots & & \vdots \\
D \xi^{(M_t-1)} & D (\xi^{(M_t-1)})^2 & \ldots & D (\xi^{(M_t-1)} )^{2^{R-1} }  \\
 & \vdots & & \vdots \\
 & \vdots & & \vdots \\
D^{\nu} \xi^{(M_t-1)} & D^{\nu} (\xi^{(M_t-1)})^2 & \ldots & D^{\nu} (\xi^{(M_t-1)} )^{2^{R-1} }  \\
\end{array}
\right]
\end{align}
Define the polynomial
\begin{equation}
\label{eq:Fdef_trellis}
f(x) = \sum_{l=0}^{R-1}u_l (D)  x^{2^{l}},
\end{equation}
where $\{u_l (D) \}_{l=0}^{R-1}\in \base[D]$. Then from
(\ref{eqn:Grepr_trellis}) with $\xi=D^{(\nu+1)(2^{R}-1)}$ we have,
\begin{align*}
\tilde{{\bf G}}^t (D) {\bf u}(D) & = \left[ \begin{array}{c}
f(1) \\
f(\xi)\\
\vdots \\
f(\xi^{(M_t-1)})\\
D f(1) \\
\vdots \\
D^{\nu} f(\xi^{(M_t-1)})
\end{array}
\right]
\end{align*}
The proof now that the left null space of $\Omega\left( \tilde{{\bf
G}}^t (D) {\bf u}(D) \right)$ over $\base$ is of dimension at most
$d(\nu+1)$ is the same as the proof of Theorem \ref{rate2} by choosing
$T$ such that,
\begin{align*}
T & \geq  (2^R-1)\nu +  (2^R-1)(\nu+1) \left( (M_t-2) (2^{R}-1)+R \right).
\end{align*}
Therefore, given the result of Theorem \ref{rate2}, which is proved in
Section \ref{subsec:general}, we can prove the rank guarantees of the
convolutional codes.

\section{Rate Guarantees}
\label{sec:Rate}
In this section we will give background needed for construction of
binary codes $\mathcal{K}_{\nu,d}$ with properties given in Definition
\ref{suitable_isi}. We start in Section \ref{subsec:CodeStr} with a
representation of $\mathcal{K}_{\nu,d}$ in terms of polynomials over
$\mathbb{F}_{2^T}$ which will be useful in proving the construction of
binary codes $\mathcal{K}_{\nu,d}$. In Section \ref{subsec:def} we
will list some definitions which we will use in proving rank
guarantees in Section \ref{sec:Rank}.  Note that these definitions are
not required for constructing $\mathcal{K}_{\nu,M_t}$, {\em i.e.,}
maximal rank sets, for which the proof is much simpler as seen in
Section \ref{subsec:maximal}. Finally in Section \ref{subsec:rate} we
will show that $|\mathcal{K}_{\nu,d}|\geq 2^{RT-\nu M_t}$, where
$R=M_t-d+1$. The rank properties of $\mathcal{K}_{\nu,d}$ are given in
Section \ref{sec:Rank}.

\subsection{Polynomial representation}
\label{subsec:CodeStr}
Given a rate $R$, we define the linearized polynomial
\begin{equation}
\label{eq:Fdef}
f(x) = \sum_{l=0}^{R-1}f_l x^{2^{l}},
\end{equation}
where $\{f_l\}_{l=0}^{R-1}\in \ext$. To develop the binary matrices
with structure given in (\ref{eq:Bdef}), we define
$\cbf_f\in\mathbb{F}_{2^T}^{M_t}$
\begin{align}
\mathbf{c}_{f} & = \left[\begin{array}{cccc} f(1) & 
f(\xi) & \ldots & f(\xi^{(M_t-1)})
\end{array} \right]^t. 
\end{align}
where $\xi=\alpha^{(2^{R}-1)(\nu+1)}$, and $\alpha$ is a primitive
element of $\ext$. Let $\mathbf{f}^{(0)}(\xi^i)$ and
$\mathbf{f}^{(k)}(\xi^i)$ be the representations of $f(\xi^i)$ and
$\alpha^k f(\xi^i)$ in the basis
$\{\alpha^0,\alpha^1,\ldots,\alpha^{T-1} \}$ respectively {\em i.e.},
$\mathbf{f}^{(0)}(\xi^i), \mathbf{f}^{(k)}(\xi^i) \in \base^{1\times
T}$ . We obtain a matrix representation $\Cbf_f\in\base^{M_t\times T}$
of $\mathbf{c}_f$ as,
\begin{align}
\mathbf{C}_{f} & =  
\left[\begin{array}{cccc} 
\mathbf{f}^{(0)t}(1) & \mathbf{f}^{(0)t}(\xi)  & \ldots & \mathbf{f}^{(0)t}(\xi^{(M_t-1)})
 \end{array} \right]^t. 
\end{align}
Now, in order to get the structure required in (\ref{eq:Bdef}), we
need to study the requirements of $f$ so that the last $\nu$ elements
in $\mathbf{C}_{f}$ are $0$ for all the $M_t$ rows. Note that the
$j^{th}$ row of $\mathbf{C}_{f}$ is given by the binary expansion of
$f(\xi^{j-1})\in\ext$ in terms of the basis
$\{\alpha^0,\alpha,\ldots,\alpha^{T-1}\}$, where $\alpha$ is a
primitive element of $\ext$. The coefficients in this basis expansion
can be obtained using the trace operator described below for
completeness\footnote{More background can be found in standard
textbooks on finite fields
\cite{lidl,mceliece}.}.

Consider an extension field $\ext$ of the base field $\base$. If
$\alpha\in\ext$ is a primitive element of $\ext$ then
$(\alpha^0,\alpha^1,\ldots,\alpha^{T-1})$ form a basis of $\ext$ over
$\base$ and any element $\beta\in \ext$ can be uniquely represented in
the form,
\begin{align*}
\beta=\beta_0 \alpha^0 +\beta_1 \alpha^1 +\ldots \beta_{T-1}
\alpha^{T-1} \quad {\rm with}\ \beta_i \in \base, \ {\rm for }\ 0 \leq
i \leq (T-1).
\end{align*}
To solve for the coefficients $\beta_i$ we will use the trace function
and trace dual bases. Note that for any element $\beta\in\ext$ the
trace of the element $\beta$ relative to the base field $\base$ is
defined as,
\begin{align*}
\tr \left( \beta \right) & = \beta+\beta^2+\beta^{2^2}+\ldots+\beta^{2^{T-1}}.
\end{align*}
Given that $\beta,\tilde{\beta} \in \ext$ the trace function satisfies
the following properties,
\begin{itemize}
\item $\tr(\beta) \in \base$.
\item $\tr (\beta+\tilde{\beta}) = \tr(\beta)+\tr(\tilde{\beta})$.
\item $\tr (\lambda \beta) = \lambda \tr(\beta)$, if $\lambda\in \base$.
\end{itemize}

Also given the basis $(\alpha^0,\alpha^1,\ldots,\alpha^{T-1})$ the
corresponding trace dual basis
$(\theta_0,\theta_1,\ldots,\theta_{T-1})$ is defined to be the unique
set of elements which satisfy the following relation for $0\leq
i,j\leq(T-1)$,
\begin{align*}
\tr\left( \theta_i \alpha^j \right) & = \begin{cases}
0 & {\rm for}\ i\neq j \\
1 & {\rm for}\ i= j 
\end{cases}
\end{align*}
The fact that the trace dual basis exists and is unique can be found
in standard references such as \cite{lidl,mceliece}. Therefore given
$\beta \in \ext$, we can find $\beta_i$ by using the properties of the
trace function and noting that,
\begin{align*}
\tr \left( \theta_i \beta \right) & = \tr \left(  \theta_i \sum_{j=0}^{T-1} \beta_j \alpha^j \right) 
 = \sum _{j=0}^{T-1} \tr \left(  \theta_i \beta_j \alpha^j \right) 
 = \sum _{j=0}^{T-1} \beta_j \tr \left(  \theta_i \alpha^j \right) 
 = \beta_i
\end{align*}
where the last equality follows from the definition of the trace dual
basis.
Therefore binary matrix $\mathbf{B}$ given in (\ref{eq:Bdef}) can be
represented in terms of the set $\mathcal{S}$ defined as
\begin{align}
  \mathcal{S}  & = \left\{ f : f \in \mathbb{F}_{2^T}^{R} , \tr \left( \theta_i
  f(\xi^j) \right)= 0 \,\, \forall i \in \{T-\nu,\ldots,T-1\} \, \, {\rm
  and} \, \, \forall j \in \{1,\ldots,M_t\} \right\}
\label{eq:Sdef}
\end{align}
Associate to $f\in\mathcal{S}$ the codeword vector $\mathbf{u}_{f}\in
\ext^{(\nu+1) M_t \times 1} $ given by,
\begin{align}
\mathbf{u}_{f} & = \left[\begin{array}{cccc} \mathbf{c}^t_{f} & \alpha \mathbf{c}^t_{f} 
& \ldots & \alpha^{\nu} \mathbf{c}^t_{f} \end{array} \right]^t
\end{align}
Associate with every such codeword $\mathbf{u}_{f}$ the
codeword matrix $\mathbf{U}_{f} \in \base^{(\nu+1) M_t \times T}$
given by the representation of each element of $\mathbf{u}_{f}$ in the
basis $\{\alpha^0,\alpha^1,\ldots,\alpha^{T-1} \}$.

Since $f \in \mathcal{S}$ we know that the last $\nu$ elements in
$\mathbf{C}_{f}$ are $0$ for all the $M_t$ rows.  Therefore we can see
that $\mathbf{f}^{(k)}(\xi^i)$ is a cyclic shift by $k$ positions of
$\mathbf{f}(\xi^i)$.  Hence, for $i \in
\{0,1,\ldots,\nu\}$ we can write,
\begin{align}
\mathbf{C}_{f}^{(i)} & = \left[\begin{array}{cccc} 
\mathbf{f}^{(i)t}(1)
&   \mathbf{f}^{(i)t}(\xi) & \ldots &  \mathbf{f}^{(i)t}(\xi^{(M_t-1)})
\end{array} \right]^t,
\end{align}
where $\mathbf{C}_{f}^{(i)}$ represents the matrix obtained by a cyclic
shift of all the rows of the matrix $\mathbf{C}_{f}$ by $i$ positions.

For transmission over an ISI channel, as seen in Section
\ref{subsec:Code_ISI}, it can be shown from equation
(\ref{eqn:model1}) that the effective binary transmitted codeword
matrix for a particular $f$ is of the form
\begin{align}
\mathbf{U}_{f} & = \left[\begin{array}{cccc} \mathbf{C}_{f}^t &  
\mathbf{C}_{f}^{(1)t} & \ldots &  \mathbf{C}_{f}^{(\nu)t} 
\end{array} \right] ^t
\end{align}
Clearly we see that $\mathcal{K}_{\nu,d}=\left\{ \mathbf{C}_{f} : f
\in \mathcal{S}, \mbox{rank}(\Ubf_f)\geq d(\nu+1)\right\}$. 
We will show in \ref{subsec:general} that indeed for $R=M_t-d+1$ that
$\mathcal{K}_{\nu,d}=\left\{ \mathbf{C}_{f} : f
\in \mathcal{S}\right\}$, {\em i.e.,} $\mbox{rank}(\Ubf_f)\geq d(\nu+1), \forall f\in \mathcal{S}$. 

\subsection{Notation and Definitions}
\label{subsec:def}
We will need the following definitions in the construction of the
basis vectors of the null space of $\Ubf_f$.
\begin{enumerate}
\item We define a set $\Gamma\subseteq \ext$ which will be used
extensively in the proof in Section \ref{sec:Rank} as,
\begin{equation}
\label{eq:GammaDef}
\Gamma = \{\gamma\in \ext : \gamma = \sum_{t=0}^{\nu} \delta_t \alpha^t, \delta_t
\in \base \}.
\end{equation}
\item
Given a binary vector $\mathbf{b}\in \base^{(\nu+1) M_t \times1}$
define $\Psi:\base^{(\nu+1) M_t
\times1} \rightarrow
\Gamma^{ 1 \times M_t} $ as,
\begin{align}
\label{eqn:DefOmega}
\Psi(\mathbf{b}) & = 
\underbrace{
\left[ 
\begin{array}{ccc}
\sum_{i=0}^{\nu} b_{i M_t+1}\alpha^i &  \ldots & \sum_{i=0}^{\nu} b_{i M_t+M_t} \alpha^i 
\end{array}
\right]
}_\mathbf{g} 
\end{align}
Note that the mapping $\Psi$ is a one-to-one mapping between
$\mathbf{b}$ and $\mathbf{g}$, due to the linear independence of
$\{\alpha^0,\alpha,\ldots,\alpha^{\nu}\}$.
\item For a given fixed $\mathbf{c}_f \in \ext^{M_t \times 1}$ define
$\mathcal{G}_f\subseteq \Gamma^{1\times M_t}$ such that,
\begin{eqnarray}
\label{eqn:definition_G}
\mathcal{G}_f = \{\mathbf{g}\in\Gamma^{1\times M_t}: \mathbf{g}
\mathbf{c}_f & = 0\}
\end{eqnarray}
\item Motivated by the mapping in (\ref{eqn:DefOmega}), for each
$\mathbf{g}^{(i)} \in \mathcal{G}_f$ we will use the following
representation:
\begin{eqnarray}
\label{eq:DeltaDef}
\mathbf{g}^{(i)} & = & \left[ \begin{array}{ccc} g_0^{(i)} & \ldots &
g_{M_t-1}^{(i)} \end{array} \right]\\ \nonumber
g_k^{(i)} & = &
\sum_{j=0}^{\nu}\delta_{k,j}^{(i)} \alpha^j \quad {\rm where} \,\,
\delta_{k,j}^{(i)}\in \base
\end{eqnarray}
\item For an element $\gamma \in \Gamma$ given by
$\gamma = \sum_{j=0}^{\nu}\delta_{j} \alpha^j$, define
\begin{eqnarray}
\label{eq:DegDef}
deg (\gamma) =  \max_j \left\{ j : \delta_{j} \neq 0 \right\}
\end{eqnarray}
\item For each $\mathbf{g}\in\mathcal{G}_f$ define,
\begin{eqnarray}
\label{eq:DegVecDef}
deg (\mathbf{g})  =  \max_k \left\{ j : \delta_{k,j} \neq 0 \right\}
\end{eqnarray}
\item For each $\mathbf{g}^{(i)}\in\mathcal{G}$ define a function $\Phi:\Gamma^{ 1 \times M_t}
\rightarrow \base^{1 \times M_t}$ by,
\begin{eqnarray}
\label{eq:PhiDef}
\Phi(\mathbf{g}^{(i)}) & = \left[ \begin{array}{cccc}
\delta_{0,0}^{(i)} & \delta_{1,0}^{(i)} & \ldots &
\delta_{M_t-1,0}^{(i)} \end{array} \right]
\end{eqnarray}
\item Given a set of elements
$\mathbf{g}^{(1)},\mathbf{g}^{(2)},\ldots, \mathbf{g}^{(d)} \in
\Gamma^{1 \times M_t}$ define,
\begin{eqnarray}
\label{eq:DsetDef}
\mathcal{D}\left(\mathbf{g}^{(1)},\mathbf{g}^{(2)},\ldots,
 \mathbf{g}^{(d)}\right) & = \left\{\mathbf{g}:\mathbf{g} =
 \sum_{i=1}^{d} \gamma_i \mathbf{g}^{(i)}, \mbox{ where for all $i$, }\gamma_i \in \Gamma,
 \,\, \gamma_i \mathbf{g}^{(i)} \in \Gamma^{1\times M_t} \right\}
\end{eqnarray}
Note that it then directly follows that,
\begin{eqnarray}
 |\mathcal{D} \left(\mathbf{g}^{(1)},\mathbf{g}^{(2)},\ldots,
  \mathbf{g}^{(d)}\right)|& \leq 2^{d(\nu+1)}. \label{eqn:sizeofD}
\end{eqnarray}
\end{enumerate}

\subsection{Set cardinality}
\label{subsec:rate}
Using the polynomial representation given in Section
\ref{subsec:CodeStr}, we can give a lower bound on the rate as
follows.
\begin{theorem}
\label{theorem_rate}
Consider $T > (\nu+1) M_t $ then a lower bound to the cardinality of
the set $\mathcal{S}$ is given by $|\mathcal{S}|\geq 2^{R T- \nu M_t}$
or lower bound to effective rate $R_{eff}=\frac{1}{T}\log|\mathcal{S}|$ is,
$R_{eff}  = R-\frac{\nu M_t}{T}$.
\end{theorem}
\begin{proof}
Let $\lambda_{\theta_i,\beta_j}$ be the mapping,
\begin{align*}
\lambda_{\theta_i,\beta_j}: \left[ \begin{array}{cccc} f_{R-1}& \ldots & f_1 & f_0
\end{array} \right]^t \mapsto \tr (\theta_i f(\beta_j)),
\end{align*}
for some $\beta_j\in\mathbb{F}_{2^T}, j=1,\ldots,M_t$.  This is
homomorphism of the $\base$-vector space $\ext^R$ into $\base$. The
cardinality of the set $\mathcal{S}$ is given by,
\begin{align*}
|\mathcal{S}| & =  \left| \displaystyle \bigcap_{i,j} {\rm
 ker}(\lambda_{\theta_i,\beta_j}) \right| \,\, i \in \{T-\nu,\ldots,T-1\}
 \, \, 
{\rm \&} \, \,
 j \in \{1,\ldots,M_t\}
\end{align*}
Note that the range space of $\lambda_{\theta_i,\beta_j}$ is the range
of the trace function, {\em i.e.,} $\{ 0,1 \}$. Noting that since $T >
(\nu+1) M_t$ and the rank of the equivalent matrix transformation of
$[\lambda_{\theta_{T-\nu},\beta_1},\ldots,\lambda_{\theta_{T-1},\beta_{M_t}}]^t$
at most $\nu M_t$ and therefore the null space is of dimension at
least $RT-\nu M_t$. Therefore, we conclude that, $|\mathcal{S}|\geq
2^{R T-
\nu M_t}$. \qed
\end{proof}
The Theorem \ref{theorem_rate} implies that we do not lose too much,
in terms of rate, by the zero padding at the end of the transmission
block. In particular it is a constant factor which does not depend on
$T$ and therefore can be made small by taking large enough $T$. Note
that this lower bound could be loose, and we may not lose as much rate
as $\frac{\nu M_t}{T}$

We still need to show that this set satisfies the rank guarantees,
which we will do next in Section \ref{sec:Rank}.

\section{Rank Guarantees}
\label{sec:Rank}
In Section \ref{sec:Rate}, see (\ref{eq:Sdef}), we have already
constructed codes (binary sets) $\mathcal{S}$ which satisfy the
structure in (\ref{eq:Bdef}) and that $|\mathcal{S}|\geq
2^{T(M_t-d+1)-\nu M_t}$. Therefore, this set $\mathcal{S}$ is a good
candidate for the construction of $\mathcal{K}_{\nu,d}$, needed for
the multilevel construction of Section \ref{subsec:Code_ISI}. In this
section we will prove that the set $\mathcal{S}$ in (\ref{eq:Sdef})
also satisfies the rank guarantees given in Definition
\ref{suitable_isi} and hence proving Lemma \ref{lem:KsetCons}.  To
illustrate the proof techniques, we will first prove the rank
guarantees for the maximal rank binary codes {\em i.e.},
$\mathcal{K}_{\nu,M_t}$ in Section \ref{subsec:maximal}. However, the
argument for arbitrary rank needs a more sophisticated argument. We
will explore the structure of the null space of $\Ubf_f$ and find a
basis for it in \ref{subsec:basis}. Using the structure of the basis
we will finally bound the cardinality and dimension of the null space
giving the required rank guarantees for $\mathcal{K}_{\nu,d}$ with
$T_{thr}= R \nu +(M_t-1)(\nu+1) (2^{R}-1)$.

\subsection{Maximal rank distance codes}
\label{subsec:maximal}
In this section we will show that that if $R=1$ then for all
$f\in\mathcal{S}, \mbox{rank}(\Ubf_f)\geq M_t(\nu+1)$.  In fact for
this case $T_{thr}=M_t(\nu+1)$ is enough.

\begin{theorem}[{\bf (Maximal rank distance codes)}]
Let $f(x)=f_0x$, as in (\ref{eq:Fdef}) with $R=1$ and $T \geq
M_t(\nu+1)$.  Then for $\mathcal{S}$ defined in (\ref{eq:Sdef}),
$\frac{1}{T}\log|\mathcal{S}|\geq 1-\frac{\nu M_t}{T} $ and $\forall
f\in\mathcal{S},
\mbox{rank}(\Ubf_f)\geq M_t(\nu+1)$ over the binary field.
\label{rate1}
\end{theorem}
\begin{proof}
The rate lower bound is directly from Theorem \ref{theorem_rate}.  We
prove the result by contradiction. Suppose that
$\mathcal{O}=\{\Ubf_f:f\in \mathcal{S}\}$ has rank distance less than
$(\nu+1)M_t$, then there exists a vector $\ubf_{f}\neq \mathbf{0}$ for
some $f \in \mathcal{S}$ such that the corresponding binary matrix
$\mathbf{U}_{f}$ has binary rank less than $(\nu+1)M_t $ (as the code
is linear). So there exists a non-trivial binary vector space
$\mathcal{B} \subseteq \mathbb{F}_2^{T}$ such that for every
$\mathbf{b} \in \mathcal{B} $,
\begin{eqnarray}
\label{eq:LeftNullSpUf}
 \mathbf{b}^t \mathbf{U}_{f}  & = &  \mathbf{0} \iff
\sum_{i=1}^{(\nu+1)M_t} b_i\Ubf_f(i,j) = 0, \,\,\, j=1,\ldots,T,
\end{eqnarray}
where $\Ubf_f(i,j)$ is the $(i,j)^th$ entry of $\Ubf_f$ and we have used
$(\cdot)^t$ to denote vector transpose. Since each row of $\Ubf_f$ is an
expansion of the rows of $\ubf_f$ in the basis
$\{\alpha^0,\alpha,\ldots,\alpha^{T-1}\}$, we can write as operations
over $\ext$,
\begin{eqnarray}
\label{eq:LeftNullSpSmallUf}
\mathbf{b}^t \mathbf{u}_{f} &=& \sum_{i=1}^{(\nu+1)M_t} b_i \ubf_f(i)
= \sum_{i=1}^{(\nu+1)M_t} b_i\sum_{j=1}^T \Ubf_f(i,j) \alpha^{j-1} =
\sum_{j=1}^T \alpha^{j-1}\left [ \sum_{i=1}^{(\nu+1)M_t} b_i \Ubf_f(i,j) \right ],
\end{eqnarray}
where we have used the basis expansion. Due to the linear independence
of $\{\alpha^0,\alpha,\ldots,\alpha^{T-1}\}$, it is clear from
(\ref{eq:LeftNullSpUf}) and (\ref{eq:LeftNullSpSmallUf})that,
\begin{eqnarray}
\label{eq:NullSpUfSmallUfEq}
\mathbf{b}^t \mathbf{U}_{f} = 0 \iff  \mathbf{b}^t \mathbf{u}_{f}  = 0.
\end{eqnarray}
Now, we suppose that for $\mathbf{b}\neq \mathbf{0}$,
\begin{align}
\mathbf{b}^t \mathbf{u}_{f} & = \sum_{i=0}^{\nu} \sum_{k=0}^{M_t-1} b_{i + k(\nu+1)} 
\alpha^i f(\alpha^{k(\nu+1)})  = \sum_{i=0}^{\nu} \sum_{k=0}^{M_t-1} b_{i + k(\nu+1)} 
\alpha^i f_0 \alpha^{k(\nu+1)} \nonumber \\
& = f_0 \left(\sum_{i=0}^{\nu} \sum_{k=0}^{M_t-1} b_{i +k(\nu+1)}
\alpha^{i+k(\nu+1)} \right) = 0 \label{eqn1}
\end{align}
Thus, for every $\mathbf{b} \in \mathcal{B}$ the element
$\left(\sum_{i=0}^{\nu} \sum_{k=0}^{M_t-1} b_{i+k(\nu+1)} \alpha^{i+k(\nu+1)}
\right)$ is a zero of $f(x)$. But we know that $\{\alpha^{i+k(\nu+1)}\}$
are linearly independent for $k\in\{0,1,\ldots,M_t-1\}$ and
$i\in\{0,1,\ldots,\nu\}$ as $T \geq (\nu+1) M_t $. Therefore there is
only one trivial solution to the equation (\ref{eqn1}) {\em i.e.},
$b_{i+k(\nu+1)}=0$ for $i=\{0,\ldots,\nu\},k=\{1,\ldots,\nu\}$. This
contradicts the fact that the null space is non-trivial since we
cannot have $\mathbf{b}\neq \mathbf{0}$ and
$\mathbf{b}\in\mathcal{B}$. Hence all matrices in $\mathcal{O}$ have
rank equal to $M_t (\nu+1)$.\qed
\end{proof}

\subsection{Minimal Basis Vectors}
\label{subsec:basis}

To prove the rank distance properties in this subsection we will show
the existence of elements which satisfy the following properties.

\begin{definition}{\bf (Properties of Minimal Basis Vectors)}
\label{def:basis}
Given a fixed nonzero vector $\mathbf{c}_f \in \ext^{M_t \times 1}$
define the associated $\mathcal{G}_f$ as in equation
(\ref{eqn:definition_G}). Then the elements $\g{1},\g{2},\ldots, \g{d}
\in \mathcal{G}_f$ are called the minimal basis vectors if they
satisfy the following properties:
\begin{enumerate}
\renewcommand\theenumi{(i)}
\item \label{prop1} For each $\g{i}$, $\exists$ $k$ such that $\delta_{k,0}^{(i)}=1$, {\em i.e.},
$\Phi(\g{i})\neq \mathbf{0}$.
\renewcommand\theenumi{(ii)}
\item \label{prop2} $\Phi(\g{1}),\ldots,\Phi(\g{d})$ are linearly independent over $\base$.
\renewcommand\theenumi{(iii)}
\item \label{prop3} For all subsets $S \subseteq \{1,\ldots,d\}$ there
do not exist $\left\{ \gamma_i : i\in S, \gamma_i \in \Gamma\,\,{\rm and}\,\,
\gamma_i \g{i} \in\Gamma^{1\times M_t} \right\}$, such that,
\begin{align*}
deg(\sum_{i\in S} \gamma_i \g{i}) & < \max_{i \in S} deg(\gamma_i \g{i}) 
\end{align*}
\renewcommand\theenumi{(iv)}
\item  \label{prop4} We have,
\begin{equation}
\label{eq:GfDsetEq}
\mathcal{G}_f  =  \mathcal{D}\left(\g{1},\g{2},\ldots, \g{d}\right),
\end{equation}
where $\mathcal{D}(\cdot,\ldots,\cdot)$ is defined as in (\ref{eq:DsetDef}) as,
\begin{eqnarray}
\label{eq:DsetDefRep}
\mathcal{D}\left(\mathbf{g}^{(1)},\mathbf{g}^{(2)},\ldots,
 \mathbf{g}^{(d)}\right) & = \left\{\mathbf{g}:\mathbf{g} =
 \sum_{i=1}^{d} \gamma_i \mathbf{g}^{(i)}, \mbox{ where for all $i$, }\gamma_i \in \Gamma,
 \,\, \gamma_i \mathbf{g}^{(i)} \in \Gamma^{1\times M_t} \right\}
\end{eqnarray}
\end{enumerate}
\end{definition}

To prove the existence of such minimal basis vectors, we need the
following lemmas. We state the lemma \ref{lemma:Prop3_Generalized}
required in the proofs and then prove it in the appendix.

\begin{lemma}
\label{lemma:Prop3_Generalized}
Assume there exist $p$ elements $\g{1},\ldots,\g{p}\in\mathcal{G}_f$
which do not satisfy property \ref{prop3} {\em i.e.}, for some subset
$S \subseteq\{1,\ldots,p\}$ there exist $\left\{ \gamma_i : i\in S,
\gamma_i \in \Gamma\,\,{\rm and}\,\, \gamma_i \g{i}
\in\Gamma^{1\times M_t} \right\}$ such that,
\begin{align*}
deg(\sum_{i\in S} \gamma_i \g{i}) & < \max_{i \in S} deg(\gamma_i \g{i}).
\end{align*}
Then there exists a set $S' \subseteq S $ and $k \in S$, $k \notin S'$
such that,
\begin{align*}
deg\left( \mathbf{g}^{(k)} + \sum_{i\in S'} \gamma_i
\mathbf{g}^{(i)} \right) & < deg\left(\mathbf{g}^{(k)}\right)
\end{align*}
and
\begin{align*}
deg\left( \gamma_i \mathbf{g}^{(i)} \right) & \leq
deg\left(\mathbf{g}^{(k)}\right) \quad \forall i \in S'
\end{align*}
where by definition we have that $\gamma_i \g{i} \in \Gamma^{1\times
M_t}$ for all $i\in S'$.
\end{lemma}

\begin{lemma}
\label{lemma:tilde_operation}
If there exist $p$ elements $\g{1},\ldots,\g{p}\in\mathcal{G}_f$
satisfying \ref{prop1}, \ref{prop2} and \ref{prop3} in Definition
\ref{def:basis} but not satisfying \ref{prop4} then it is possible to
form $\tg{1},\ldots,\tg{p},\tg{p+1}$ satisfying \ref{prop1},
\ref{prop2} and,
\begin{align}
\label{eq:tilde_containment}
\mathcal{D}(\g{1},\ldots,\g{p}) & \subset  \mathcal{D}(\tg{1},\ldots,\tg{p},\tg{p+1})
\end{align}
\end{lemma}
\begin{proof}
Since we have $\g{1},\ldots,\g{p}$ satisfying \ref{prop1}, \ref{prop2}
and \ref{prop3} but not satisfying \ref{prop4} there exists
$\g{p+1}\in\mathcal{G}_f$ such that $\g{p+1}\notin
\mathcal{D}(\g{1},\ldots,\g{p})$. If $\Phi(\g{p+1})=\mathbf{0}$, then
clearly we can write $\g{p+1}=\alpha^t\breve{\gbf}^{(p+1)}$, where
$\Phi(\breve{\gbf}^{(p+1)})\neq\mathbf{0}$, since we are only taking
out the common $\alpha^{(\cdot)}$ factor out of $\g{p+1}$. Note that
clearly $\breve{\gbf}^{(p+1)}\in\mathcal{G}_f$ and since
$\g{1},\ldots,\g{p}$ satisfy \ref{prop3}, we can show that
$\breve{\gbf}^{(p+1)} \notin
\mathcal{D}(\g{1},\ldots,\g{p}).$\footnote{ Assume that
$\breve{\gbf}^{(p+1)} \in \mathcal{D}(\g{1},\ldots,\g{p})$ but
$\g{p+1}\notin \mathcal{D}(\g{1},\ldots,\g{p})$. Since
$\g{p+1}\in\mathcal{G}_f$ we have $\g{p+1}\in \Gamma^{1\times
M_t}$. The set $\mathcal{D}(\g{1},\ldots,\g{p})$ contains all
combinations of $\gamma_i \g{i}$ such that $\gamma_i
\g{i}\in\Gamma^{1\times M_t}$.  The only way this is possible is if
for some set of $\{\gamma_i\}$, $\sum \alpha^t(\gamma_i \g{i}) \in
\Gamma^{1 \times M_t}$ but $\alpha^t\gamma_k\g{k} \notin
\Gamma^{1\times M_t}$ for some $k$. This implies that,
\begin{align*}
deg\left(\alpha^t \gamma_{k}\g{k} + \sum \alpha^t(\gamma_i \g{i})\right) & < deg\left( \alpha^t\gamma_k\g{k}\right).
\end{align*}
Since $\g{1},\ldots,\g{p}$ satisfy \ref{prop3} this is not
possible. Therefore, we can always choose $\g{p+1}\in\mathcal{G}_f$
such that $\g{p+1}\notin \mathcal{D}(\g{1},\ldots,\g{p})$ such that
$\Phi(\g{p+1})\neq\mathbf{0}$.}

If $\Phi(\g{p+1})$ is linearly independent of
$\Phi(\g{1}),\ldots,\Phi(\g{p})$ then $(\g{1},\ldots,\g{p+1})$ satisfy
\ref{prop1} and \ref{prop2} and (\ref{eq:tilde_containment}) follows
directly by choosing $\tg{i}=\g{i}, i=1,\ldots p+1$. 

If $\Phi(\g{1}),\ldots,\Phi(\g{p+1})$ are not linearly independent
then,
\begin{align}
w_1 \Phi(\g{1}) + w_2 \Phi(\g{2}) +\ldots w_{p+1}\Phi(\g{p+1}) & = 0
\end{align}
for $w_1,\ldots,w_{p+1} \in \base$ and not all equal to zero. Let
$\g{k}$ be such that $w_k \neq 0 $ and 
\begin{align}
deg(\g{k}) & \geq  deg(\g{i}) \quad \forall i, \, such\, \, that\, \, w_i=1
\end{align}
Since $w_1,\ldots,w_{p+1} \in \base$, we see that, $w_1 \Phi(\g{1}) +
w_2 \Phi(\g{2}) +\ldots w_{p+1}\Phi(\g{p+1})=\Phi(w_1
\g{1}+\ldots+w_{p+1} \g{p+1})=0$.  Therefore, there is a common
$\alpha^{(\cdot)}$ factor in $w_1 \g{1}+\ldots+w_{p+1} \g{p+1}$ {\em
i.e.,} there is $\tg{k}$ and $t$ such that,
\begin{displaymath}
\left(w_1 \g{1}+\ldots+w_k\g{k}+\ldots+w_{p+1} \g{p+1}  \right) = \alpha^{t}\tg{k},
\end{displaymath}
where $t$ is chosen to be the minimum value such that
$\Phi(\tg{k})\neq\mathbf{0}$.  Using this we can define,
\begin{eqnarray}
\label{eq:IterDef}
\tg{k} & = & \alpha^{-t} \left(w_1 \g{1}+\ldots+\g{k}+\ldots+w_{p+1} \g{p+1}  \right) \\ \nonumber
\tg{i} & = & \g{i} \quad \forall i \neq k,
\end{eqnarray}
where we have used the fact that $w_k=1$.  Note that
\begin{align}
\label{eq:TgDegBnd}
deg(\tg{k}) & \leq deg(\g{k})-t
\end{align}
Clearly \ref{prop1} is satisfied for $\tg{1},\ldots,\tg{p+1}$. Moreover,
$\tg{1},\ldots,\tg{p+1}\in \mathcal{G}_f$ since
$\g{1},\ldots,\g{p+1}\in\mathcal{G}_f$. We will now show that,
\begin{align}
\mathcal{D}(\g{1},\ldots,\g{p+1}) & \subset  \mathcal{D}(\tg{1},\ldots,\tg{p+1})
\end{align}
Let $\mathbf{g}\in \mathcal{D}(\g{1},\ldots,\g{p+1})$, {\em i.e.}, 
\begin{align}
\mathbf{g} & =  \gamma_1 \g{1} + \ldots + \gamma_{p+1} \g{p+1}
\end{align}
such that $\gamma_i \g{i} \in \Gamma^{1\times M_t}$. Note the important fact
that since $\gamma_i \g{i} \in \Gamma^{1\times M_t}$ we have that
\begin{equation}
\label{eq:Lem12DegBnd}
deg(\gamma_i)+ deg(\g{i}) \leq \nu, 
\end{equation}
where we have used the definitions given in (\ref{eq:DegDef}) and
(\ref{eq:DegVecDef}). Now consider,
\begin{align}
\label{eq:TgDegBnd2}
\tilde{\gamma}_k & =  \gamma_k \alpha^t \\
\tilde{\gamma}_i & =  w_i \gamma_k+\gamma_i \quad \forall i \neq k \nonumber
\end{align}
Then,
\begin{align*}
\tilde{\gamma}_1 \tg{1}+ \ldots+ \tilde{\gamma}_{p+1} \tg{p+1} & =
\gamma_k \alpha^{t} \left[ \alpha^{-t}(w_1
\g{1}+\ldots+\g{k}+\ldots+w_{p+1} \g{p+1} ) \right]+\sum_{i\neq k
}(w_i\gamma_k+\gamma_i)\g{i} \\ & = \gamma_k \g{k} + \sum_{i\neq k}
\left( w_i \gamma_k \g{i}\right)+ \sum_{i\neq k} \left(\gamma_i \g{i} +
w_i \gamma_k \g{i} \right) \\ & = \gamma_1 \g{1} + \ldots +
\gamma_{p+1} \g{p+1}
\end{align*}
where the last step follows as the field has characteristic $2$. The
only thing to verify is that $\tilde{\gamma}_i\in\Gamma$ and $
\tilde{\gamma}_i\tg{i}\in \Gamma^{1 \times M_t}$. Trivially, $\tilde{\gamma}_i
\in \Gamma$ and $\tilde{\gamma}_i\tg{i}\in \Gamma^{1 \times M_t}$ for all
$i\neq k$. Also note that since,
\begin{align*}
deg(\tilde{\gamma}_k)+deg(\tg{k})& \stackrel{(a)}{=} deg(\gamma_k)+t+deg(\tg{k}) \\ &
\stackrel{(b)}{\leq} deg(\gamma_k)+deg(\g{k}) \stackrel{(c)}{\leq} \nu,
\end{align*}
where $(a)$ follows due to (\ref{eq:TgDegBnd2}), $(b)$ follows from
(\ref{eq:TgDegBnd}) and $(c)$ follows from (\ref{eq:Lem12DegBnd}).  Therefore,
$\tilde{\gamma}_k \in \Gamma$ and $\tilde{\gamma}_k \tg{k} \in
\Gamma^{1\times M_t}$. Hence $\forall \mathbf{g} \in
\mathcal{D}(\g{1},\ldots,\g{p+1})$, $\mathbf{g} \in
\mathcal{D}(\tg{1},\ldots,\tg{p+1})$. Therefore,
\begin{align}
\label{eq:NestProp}
\mathcal{D}(\g{1},\ldots,\g{p+1}) & \subset  \mathcal{D}(\tg{1},\ldots,\tg{p+1})
\end{align}
Also, since $t\geq 1$,
\begin{align}
deg(\tg{k}) & <    deg(\g{k}) 
\end{align}
and $deg(\tg{i})=deg(\g{i})$ $\forall i \neq k$.  Therefore for the
new set $\{\tg{i}\}_{i=1}^{p+1}$, the degree is smaller than or equal
to that of the previous set $\{\g{i}\}_{i=1}^{p+1}$. Therefore, since
we are reducing the degree of atleast one element and the maximal
degree of the set is bounded above by $\nu$, if we iterate this step,
the process will terminate.  We utilize this idea in the
following. Now we check if $\Phi(\tg{1}),\ldots,\Phi(\tg{p+1})$ are
linearly independent. If not, we continue the process defined in
(\ref{eq:IterDef}) till we obtain $\tg{1},\ldots,\tg{p+1}$ such that
$\Phi(\tg{p}),\ldots,\Phi(\tg{p+1})$ are linearly independent or
$deg(\tg{1})=\ldots=deg(\tg{p+1})=0$. If the former occurs, we have
obtained the required set $\{\tg{i}\}$. If the latter occurs, and if
$\Phi(\tg{p}),\ldots,\Phi(\tg{p+1})$ are linearly independent, again
we are done. Now, if the latter occurs, {\em i.e.,}
$deg(\tg{1})=\ldots=deg(\tg{p+1})=0$ and $\Phi(\tg{1}),\ldots,
\Phi(\tg{p+1})$ are linearly dependent, then since the degrees are
equal to zero we just take the set of independent $\tg{i}$.  We know
that using these sets of vectors we can satisfy \ref{prop1} and
\ref{prop2}. Note that $\mathcal{D}(\g{1},\ldots,\g{p+1})$ cannot be
equal to the set $\mathcal{G}_f$ without the elements
$(\g{1},\ldots,\g{p+1})$ satisfying properties \ref{prop1},
\ref{prop2}\footnote{\label{foot}If property \ref{prop2} is not satisfied,
$\Phi(\g{1}),\ldots,\Phi(\g{p+1})$ are not linearly independent {\em
i.e.},
\begin{align}
w_1 \Phi(\g{1}) + w_2 \Phi(\g{2}) +\ldots w_{p+1}\Phi(\g{p+1}) & = 0
\end{align}
for $w_1,\ldots,w_{p+1} \in \base$ and not all equal to zero. Since
$\sum_i w_i\Phi(\g{i})=\Phi\left (\sum_i w_i(\g{i})\right )=0$, there
is a common $\alpha^{(\cdot)}$ factor in $w_1 \g{1}+\ldots+w_{p+1}
\g{p+1}$ and the element
$\gbf=\sum_i\underbrace{\alpha^{-1}w_i}_{\gamma_i}\g{i}$ is contained
in $\mathcal{G}_f$ but not in $\mathcal{D}(\g{1},\ldots,\g{p+1})$,
since $\gamma_i\notin \Gamma$.}.  Therefore, using this iterative
process we can construct the required set $\{\tg{i}\}$ since in
(\ref{eq:NestProp}) we have already shown that the nesting property
needed in (\ref{eq:tilde_containment}) is satisfied.  \qed
\end{proof}

Note that in lemma \ref{lemma:tilde_operation}
$\mathcal{D}(\g{1},\ldots,\g{p})$ is a proper subset of
$\mathcal{D}(\tg{1},\ldots,\tg{t})$ as the element $\g{p+1}$ is not
contained in $\mathcal{D}(\g{1},\ldots,\g{p})$. 

\begin{lemma}
\label{lemma:hat_operation}
If there exist $\tg{1},\ldots,\tg{p+1}\in\mathcal{G}_f$ satisfying
\ref{prop1} and \ref{prop2} but not satisfying \ref{prop3} in
Definition \ref{def:basis} it is possible to construct
$\hg{1},\ldots,\hg{p+1}$ satisfying \ref{prop1}, \ref{prop2} and
\ref{prop3} in Definition \ref{def:basis} and
\begin{align}
\label{eq:hat_containment}
\mathcal{D}(\tg{1},\ldots,\tg{p+1}) & \subset \mathcal{D}(\hg{1},\ldots,\hg{p+1}) 
\end{align}
\end{lemma}
\begin{proof}
Given $\tg{1},\ldots,\tg{p+1}\in\mathcal{G}_f$ satisfying \ref{prop1}
and \ref{prop2} but not satisfying \ref{prop3} in Definition
\ref{def:basis}. From lemma \ref{lemma:Prop3_Generalized} we conclude that
there exists a set $S\subset\{1,\ldots,d\}$ and $k\notin S$ and
$\{\gamma_i\})_{i\in S}$ where $\gamma_i\in\Gamma, i\in S$, such that,
\begin{eqnarray}
\label{eq:Prop3NSa}
deg\left(\tg{k}+\sum_{i\in S} \gamma_i \tg{i} \right) & < & deg(\tg{k})
\end{eqnarray}
{\em and} we also have 
\begin{eqnarray}
\label{eq:Prop3NSb}
deg(\gamma_i \tg{i}) & \leq & deg(\tg{k}) \quad \forall i\in S.
\end{eqnarray}
Therefore we have $deg(\tg{k})\geq deg(\tg{i})$ for all $i\in
S$. Define,
\begin{eqnarray}
\label{eq:HatGdef}
\hg{k} & = & \tg{k}+\sum_{i\in S} \gamma_i \tg{i} \\ \nonumber
\hg{i} & = & \tg{i} \quad \forall i \neq k
\end{eqnarray}
First we show that property \ref{prop1} is satisfied by
$\hg{1},\ldots,\hg{p+1}$. This can be easily seen from the following.
We already know that $\Phi(\hg{i})\neq \mathbf{0}$ since $\hg{i} =
\tg{i}$. Moreover, $\{\Phi(\hg{i})\}_{i\neq k}$ are linearly
independent since we know that $\{\tg{i}\}$ satisfy \ref{prop1} and
\ref{prop2} of Definition \ref{def:basis}. Now, let
$\gamma_i=\sum_{b=0}^{\nu} \delta_b^{(i)} \alpha^b$ where
$\delta_b^{(i)} \in \base$. Then we have,
\begin{eqnarray}
\label{eq:PhiHatGk}
\Phi(\hg{k}) & = &\Phi(\tg{k}+\sum_{i\in S} \sum_{b=0}^{\nu}
\delta_b^{(i)} \alpha^b \tg{i}) \\ \nonumber
& = &\Phi(\tg{k})+\sum_{i\in
S}\left( \delta_0^{(i)} \Phi(\tg{i})\right) +\sum_{i\in S}
\sum_{b=1}^{\nu}\delta_b^{(i)} \Phi( \alpha^b \tg{i}) \\ \nonumber
& = & \Phi(\tg{k})+\sum_{i\in S} \delta_0^{(i)} \Phi(\tg{i})
\end{eqnarray}
Note that since $\delta_0 \in \base$ and
$\Phi(\tg{1}),\ldots,\Phi(\tg{p+1})$ were independent to begin with,
and since $\{\tg{i}\}$ satisfy \ref{prop1}, we see that the above
implies that $\Phi(\hg{k})\neq \mathbf{0}$ and hence
$\hg{1},\ldots,\hg{p+1}$ satisfy property \ref{prop1} of Definition
\ref{def:basis}.

Now suppose that $\Phi(\hg{k})$ is linearly dependent on
$\{\Phi(\hg{i})\}_{i\neq k}$, since $\hg{i}=\tg{i}, i\neq k$ we can
write,
\begin{eqnarray*}
\Phi(\hg{k}) = \sum_{i\neq k} \theta_i \Phi(\hg{i}) = \sum_{i\neq k}
\theta_i \Phi(\tg{i}),
\end{eqnarray*}
for $\theta_i\in \base$, where since $\Phi(\hg{k})\neq \mathbf{0}$ we
have that $\{\theta_i\}$ is not all zero. Due to (\ref{eq:PhiHatGk})
this implies that,
\begin{eqnarray*}
\Phi(\tg{k})+\sum_{i\in S} \delta_0^{(i)} \Phi(\tg{i}) = \sum_{i\neq
k} \theta_i \Phi(\tg{i}),
\end{eqnarray*}
which means that 
\begin{eqnarray*}
\Phi(\tg{k})=\sum_{i\in S} \delta_0^{(i)} \Phi(\tg{i}) + \sum_{i\neq
k} \theta_i \Phi(\hg{i}),
\end{eqnarray*}
which contradicts the linear independence of
$\Phi(\tg{1}),\ldots,\Phi(\tg{p+1})$ since $k\notin S$.  This implies
that $\Phi(\hg{k})$ is linearly independent of
$\Phi(\hg{i})=\Phi(\tg{i})$ for all $i\neq k$. Therefore
$\hg{1},\ldots,\hg{p+1}$ satisfy \ref{prop1} and \ref{prop2} of
Definition \ref{def:basis}.

To show (\ref{eq:hat_containment}) let
$\mathbf{\tilde{g}}\in\mathcal{D}(\tg{1},\ldots,\tg{p+1})$, {\em
i.e.},
\begin{align}
\label{eq:TildeGexp}
\mathbf{\tilde{g}} & =
\tilde{\gamma}_1\tg{1}+\ldots+\tilde{\gamma}_{p+1} \tg{p+1}.
\end{align} 
Choose $\hat{\gamma}_i = \tilde{\gamma}_i.\,\, i\notin S$, and
$\hat{\gamma}_i = \tilde{\gamma}_i+\gamma_i \tilde{\gamma}_k$ for all
$i\in S$. Note that $\{\gamma_i\}_{i\in S}$ is defined in
(\ref{eq:Prop3NSa}). Therefore, since $k\notin S$, we have
\begin{align*}
\hat{\gamma}_1 \hg{1}+ \ldots+\hat{\gamma}_{p+1} \hg{p+1} & =
\tilde{\gamma}_k \left(\tg{k}+\sum_{i \in S} \gamma_i
\tg{i}\right)+\sum_{i\in S}\left(\tilde{\gamma}_i+\gamma_i
\tilde{\gamma}_k\right)\tg{i} + \sum_{i \notin S, i\neq k}
\tilde{\gamma}_i \tg{i}\\ & = \tilde{\gamma}_k \tg{k} + \left(\sum_{i\in S}
\tilde{\gamma}_k \gamma_i \tg{i}\right) + \left(\sum_{i\in S}
\tilde{\gamma}_i\tg{i}\right) + \left(\sum_{i\in S }\gamma_i
\tilde{\gamma}_k \tg{i} \right) + \left(\sum_{i \notin S, i\neq k}
\tilde{\gamma}_i\tg{i}\right)\\ & =
\tilde{\gamma}_k\tg{k}+\left(\sum_{i\neq k}
\tilde{\gamma}_i\tg{i}\right)
\end{align*} 
where the last step follows as the characteristic of the field is
$2$. We still need to show that $\hat{\gamma}_i \in \Gamma$ and
$\hat{\gamma}_i \hg{i} \in \Gamma^{1\times M_t}$. Note that due to
(\ref{eq:Prop3NSb}), we have
\begin{align}
\label{eq:DegBndOnTgk}
deg(\tg{i})+deg(\tilde{\gamma}_i) & \leq  deg(\tg{k}) \quad \forall i\in S 
\end{align}
Also, since we have (\ref{eq:TildeGexp}), we know that
$\tilde{\gamma}_{i}\in\Gamma, \,\forall i$ and $\tilde{\gamma}_{i}
\tg{i}\in\Gamma^{1\times M_t}, \forall i$, hence
\begin{align}
\label{eq:DegBndOnTG}
deg(\tg{i})+deg(\tilde{\gamma}_i) & \leq  \nu, \,\,\forall i.
\end{align}
Since we have $\hat{\gamma}_i = \tilde{\gamma}_i.\,\, i\notin S$, and
from (\ref{eq:HatGdef}), we know that $\hg{i}=\tg{i},i\notin S, i\neq
k$, we see that $\hat{\gamma}_i\in\Gamma$ and
$\hat{\gamma}_i\hg{i}\in\Gamma^{1\times M_t}$, for all $i\notin S,
i\neq k$. 

Now, for $i\in S$, we have that $\hat{\gamma}_i =
\tilde{\gamma}_i+\gamma_i \tilde{\gamma}_k$ and
$\hat{\gamma}_i\hg{i}=\hat{\gamma}_i\tg{i}$. Therefore, for $i\in S$ we have,
\begin{eqnarray}
\label{eq:BndDegHatGam}
deg(\hat{\gamma}_i) = \max\{deg(\tilde{\gamma}_i),deg(\gamma_i)+deg(\tilde{\gamma}_k) \}.
\end{eqnarray}
We know from (\ref{eq:TildeGexp}) that $deg(\tilde{\gamma}_i)\leq \nu$. Also, 
from (\ref{eq:DegBndOnTG}) and (\ref{eq:DegBndOnTgk}), we see that for $i\in S$,
\begin{eqnarray}
\label{eq:UsefulDegBnd}
deg(\gamma_i)+deg(\tilde{\gamma}_k)+deg(\tg{i})\leq \nu,
\end{eqnarray}
which implies that $deg(\gamma_i)+deg(\tilde{\gamma}_k)\leq \nu$ and
hence from (\ref{eq:BndDegHatGam}) $deg(\hat{\gamma}_i)\leq \nu$, {\em
i.e.,} $\hat{\gamma}_i\in \Gamma,i\in S$. Also,
$deg(\hat{\gamma}_i\hg{i})\leq deg(\hat{\gamma}_i)+deg(\tg{i}),i\in
S$. Therefore, using (\ref{eq:BndDegHatGam}), we see that,
\begin{eqnarray}
\label{eq:GamHatGhat}
deg(\hat{\gamma}_i\hg{i}) \leq
\max\{deg(\tilde{\gamma}_i\tg{i}),deg(\gamma_i)+deg(\tilde{\gamma}_k)+deg(\tg{i})\}
\end{eqnarray}
We know from (\ref{eq:TildeGexp}) that
$deg(\tilde{\gamma}_i\tg{i})\leq \nu$. Now, with this and from
(\ref{eq:UsefulDegBnd}), we see that $deg(\hat{\gamma}_i\hg{i})\leq
\nu$ and hence $\hat{\gamma}_i\hg{i}\in\Gamma^{1\times M_t}$. Hence
for $i\in S$ as well we have $deg(\hat{\gamma}_i)\leq \nu$ and
$\hat{\gamma}_i\hg{i}\in\Gamma^{1\times M_t}$.

Now for $i=k$, it is clear that
$\hat{\gamma}_k=\tilde{\gamma}_k\in\Gamma$. Now we need to show that
$\hat{\gamma}_k\hg{k}\in\Gamma^{1\times M_t}$.  Therefore, we need to
show that
\begin{eqnarray*}
deg\left( \tilde{\gamma}_k \left( \tg{k}+\sum_{i\in S} \gamma_i \tg{i}
\right) \right) \leq \nu
\end{eqnarray*}
Note that $ deg (\tilde{\gamma}_k \tg{k} ) \leq \nu$ follows directly
from (\ref{eq:DegBndOnTG}). Also, from (\ref{eq:Prop3NSb}) we have
that,
\begin{eqnarray*}
deg(\gamma_i)+deg(\tg{i}) & \leq deg(\tg{k})
\end{eqnarray*}
Therefore, 
\begin{align*}
deg(\tilde{\gamma}_k \gamma_i \tg{i} ) \leq
deg(\tilde{\gamma}_k)+deg(\gamma_i)+deg(\tg{i}) & \leq deg(\tg{k})+deg(\tilde{\gamma}_k)
\leq \nu 
\end{align*}
From (\ref{eq:GamHatGhat}) we see that $deg(\hat{\gamma}_i\hg{i}) \leq
\nu$ and hence $\hat{\gamma}_k\hg{k}\in\Gamma^{1\times M_t}$.

Therefore,
\begin{align*}
\mathcal{D}(\tg{1},\ldots,\tg{p+1}) & \subset \mathcal{D}(\hg{1},\ldots,\hg{p+1}) 
\end{align*}
Note that the degree of one of the elements of the set
$\hg{1},\ldots,\hg{p+1}$ (specifically $\hg{k}$) is strictly less than
the degree of $\tg{k}$ and the degree of all other elements is the
same.  If $\hg{1},\ldots,\hg{p+1}$ satisfy \ref{prop3} then we terminate
otherwise we repeat the process. Note that at each iteration we
decrease the degree of one of the elements by at least $1$. Since we
started off with a finite degree we continue this process either until
the property \ref{prop3} is satisfied or all the elements have degree
0. At this point if property \ref{prop3} is not satisfied from lemma
\ref{lemma:Prop3_Generalized} we have for some
$S'\subset\{1,\ldots,p+1\}$ that\footnote{Since from Lemma
\ref{lemma:Prop3_Generalized}, if property \ref{prop3} is not
satisfied, then $deg\left( \gamma_i \mathbf{g}^{(i)} \right) \leq
deg\left(\mathbf{g}^{(k)}\right) \quad \forall i \in S'$, and hence we
see that for this case, $\gamma_i=1,i\in S'$.}
\begin{align*}
deg\left( \tg{k}+\sum_{i\in S'} \tg{i} \right) & < deg(\tg{k})=0
\end{align*}
This is possible only if,
\begin{align*}
\tg{k}+\sum_{i\in S'} \tg{i}  & = \mathbf{0}
\end{align*}
But since $\tg{i}=\Phi(\tg{i}), i\in S'\mbox{ or } i=k$ and we know
that $\{ \tg{i} \}$ satisfy property \ref{prop2} we get a
contradiction. Therefore property \ref{prop3} will be satisfied when
the degree of all the elements is 0.  Note that
$\mathcal{D}(\hg{1},\ldots,\hg{p+1})$ cannot be equal to the set
$\mathcal{G}_f$ without the elements $(\hg{1},\ldots,\hg{p+1})$
satisfying property \ref{prop3} 
\footnote{\label{foot2} If property \ref{prop3} is not satisfied, we have
that for some subset $S \subseteq\{1,\ldots,p\}$ there exist $\left\{
\gamma_i\right \}_{i\in S}$ such that,
\begin{align}
deg(\sum_{i\in S} \gamma_i \hg{i}) & < \max_{i \in S} deg(\gamma_i\hg{i})
\end{align}
Let $t=\max_{i \in S} deg(\gamma_i \g{i})$ and define
$k=\argmax_{i \in S} deg(\gamma_i\g{i})$. Note then
that the element,
\begin{align*}
\gbf & = \alpha^{\nu-t+1} \left( \sum_{i\in S} \gamma_i \hg{i} \right) 
\end{align*}
is contained in $\mathcal{G}_f$ as the elements
$(\hg{1},\ldots,\hg{p+1})$ satisfy property \ref{prop3} for the
$\{\gamma_i\}$. But,
\begin{align*}
\gbf & = \underbrace{\alpha^{\nu-t+1}\gamma_k}_\gamma \hg{k}+\sum_{i
\in S,i\neq k} \alpha^{\nu-t+1}\gamma_i \hg{i}
\end{align*}
is not contained in $\mathcal{D}(\hg{1},\ldots,\hg{p+1})$ because
$\gamma \notin \Gamma$.
}
\end{proof}
Given these two lemmas we will show that given a fixed nonzero
$\mathbf{c}_f \in \ext^{M_t \times 1}$ and the associated
$\mathcal{G}_f$ defined as in equation (\ref{eqn:definition_G}), there
exist minimal basis vectors satisfying the properties in Definition
\ref{def:basis}, reproduced in the following theorem for completeness.

\begin{theorem}{\bf (Existence of Minimal Basis Vectors)}
\label{theorem:basis}
Given a fixed nonzero $\mathbf{c}_f \in \ext^{M_t \times 1}$ define
the associated $\mathcal{G}_f$ as in equation
(\ref{eqn:definition_G}). Then there exist elements
$\g{1},\g{2},\ldots, \g{d} \in
\mathcal{G}_f$ such that they satisfy the following properties:
\begin{enumerate}
\renewcommand\theenumi{(i)}
\item For each $\g{i}$, $\exists$ $k$ such that $\delta_{k,0}^{(i)}=1$, {\em i.e.},
$\Phi(\g{i})\neq \mathbf{0}$. 
\renewcommand\theenumi{(ii)}
\item $\Phi(\g{1}),\ldots,\Phi(\g{d})$ are linearly independent over $\base$.
\renewcommand\theenumi{(iii)}
\item For all subsets $S \subseteq \{1,\ldots,d\}$ there
do not exist $\left\{ \gamma_i : i\in S, \gamma_i \in \Gamma\,\,{\rm and}\,\,
\gamma_i \g{i} \in\Gamma^{1\times M_t} \right\}$, such that,
\begin{align*}
deg(\sum_{i\in S} \gamma_i \g{i}) & < \max_{i \in S} deg(\gamma_i \g{i}) 
\end{align*}
\renewcommand\theenumi{(iv)}
\item  $\mathcal{G}_f  =  \mathcal{D}\left(\g{1},\g{2},\ldots, \g{d}\right)$
\end{enumerate}
\end{theorem}
\begin{proof}
Clearly let us assume $\mathcal{G}_f$ is not empty. Then $\exists$ a
$\g{1}\in \mathcal{G}_f$ such that $\delta_{k,0}^{(1)}=1$ for some
$k$, since otherwise in the $\g{1}$ picked we can take out
$\alpha^{(\cdot)}$ factor and still have it in
$\mathcal{G}_f$. Clearly properties \ref{prop2} and \ref{prop3} of
Definition \ref{def:basis} are satisfied trivially. If $\g{1}$
satisfies property \ref{prop4} then we are done. If not, we proceed to
build the set $\g{1},\ldots, \g{d}$. If $\g{1}$ does not satisfy
\ref{prop4} it means that $\exists$ $\g{2}\in\mathcal{G}_f$ such that
$\g{2}\neq\gamma_1 \g{1}$ for any $\gamma_1 \in \Gamma$ and $\gamma_1
\g{1} \in \Gamma^{1\times M_t}$. From Lemma \ref{lemma:tilde_operation} we can
construct either $\tg{1},\tg{2}$ (or just $\tg{1}$) such that they
satisfy \ref{prop1} and \ref{prop2} and
\begin{align}
\mathcal{D}(\g{1},\g{2}) & \subset  \mathcal{D}(\tg{1},\tg{2}).
\end{align}
If \ref{prop3}, \ref{prop4} are also satisfied, then $d=2$.

If \ref{prop3} is not satisfied by these vectors $\tg{1},\tg{2}$ we
can construct $\hg{1},\hg{2}$ from Lemma \ref{lemma:hat_operation}
which satisfy \ref{prop1}, \ref{prop2} and \ref{prop3}
\footnote{The reason we need the property \ref{prop3} is as follows. If we
take any element $\mathbf{g} \in \mathcal{G}_f$ then if
$\alpha\mathbf{g}\in\Gamma^{1 \times M_t}$, then $\alpha\mathbf{g}$ is
also in $\mathcal{G}_f$. This may not be captured in our definition of
$\mathcal{D}$ framework for the following reason. If $deg\left[ \tg{1}
+ \gamma \tg{2}\right] < deg(\tg{1})$ and $deg(\tg{1})\geq
deg(\tg{2})$, then for some $t$, $\alpha^t(\tg{1}+\gamma\tg{2}) \in
\mathcal{G}$ but $\alpha^t\tg{1} + \alpha^t \gamma \tg{2} \notin
\mathcal{D}(\tg{1},\tg{2})$ since $\alpha^t\tg{1}$ or
$\alpha^t\gamma\tg{2} \notin \Gamma^{1 \times M_t}$.}.

Now if $\hg{1},\hg{2}$ satisfy \ref{prop4} then we are done, otherwise
we again use the lemma \ref{lemma:tilde_operation} with
$\hg{1},\hg{2}$ as the input vectors.  Repeat this process until
$\hg{1},\ldots,\hg{d}$ satisfy the properties \ref{prop1},
\ref{prop2}, \ref{prop3} and \ref{prop4}. This process has to
terminate since we know that $|\mathcal{G}_f|\leq |\Gamma^{1\times
M_t}|\leq 2^{(\nu+1)M_t}$ and hence is finite.

\qed
\end{proof}
Note that from property \ref{prop2} the elements are such that
$\Phi(\g{1}),\ldots,\Phi(\g{d})$ are linearly independent only over
$\base$. The following lemma shows that as long as $T>(\nu+1) M_t$
this is sufficient to guarantee the independence of
$\g{1},\g{2},\ldots,\g{d}$ over $\ext$ as well.

\begin{lemma}
Consider elements $\g{1},\g{2},\ldots,\g{d}\in \mathcal{G}_f$ such
that $\Phi(\g{1}),\ldots,\Phi(\g{d})$ are linearly independent over
$\base$. If the size of the extension field $\ext$ is such
that $T>(\nu+1) M_t$ then these vectors are linearly independent over
$\ext$ as well.
\end{lemma}
\begin{proof}
Clearly $d\leq M_t$ otherwise the property \ref{prop2} in the theorem
\ref{theorem:basis} will be violated. Define, 
\begin{eqnarray*}
\mathbf{Q} & = \left[
\begin{array}{ccc}
\g{1} &
\ldots &
\g{d}
\end{array}
\right]^t
\end{eqnarray*}
and 
\begin{eqnarray*}
\mathbf{H} & = \left[
\begin{array}{ccc}
\Phi(\g{1})&
\ldots
\Phi(\g{d})
\end{array}
\right]^t
\end{eqnarray*}
By the linear independence of $\Phi(\g{1}),\ldots,\Phi(\g{d})$ we
conclude that $\mathbf{H}$ has full rank over $\base$. Therefore,
there exist $d$ linearly independent columns over $\base$ in
$\mathbf{H} \in \base^{d\times M_t}$. Select these $d$ columns and form
the matrix $\mathbf{\hat{H}}\in \base^{d \times d}$ which is of rank
$d$. Therefore $det(\mathbf{\hat{H}})=1$ as $det(\mathbf{\hat{H}})\in
\base$. Select these same columns in the matrix $\mathbf{Q}$ and 
form the matrix $\mathbf{\hat{Q}}\in \Gamma^{d\times d}$.  Let us look
at the determinant of $\mathbf{\hat{Q}}\in\ext$. Note that since
$\mathbf{\hat{Q}}\in \Gamma^{d\times d}$,
\begin{eqnarray*}
det(\mathbf{\hat{Q}}) & = \sum_{k=0}^{d \nu}\delta_k \alpha^k.
\end{eqnarray*}
Since $T> (\nu+1)M_t \geq (\nu+1) d$ we see the linear independence of
$1,\alpha,\ldots,\alpha^{T-1}$.  Moreover, note that since
$\delta_0=det(\mathbf{\hat{H}})\neq 0$ from above and therefore we
conclude that $det(\mathbf{\hat{Q}})\neq 0$. Hence the vectors
$\g{1},\ldots,\g{d}$ are linearly independent over $\ext$.  \qed
\end{proof}

\subsection{General Rank Distance Codes}
\label{subsec:general}
In this section we will prove the required rank guarantees for
$\mathcal{S}$ with $T_{thr}= R \nu +(M_t-1)(\nu+1) (2^{R}-1)$ and
therefore show that $\mathcal{K}_{\nu,d}$ is given by this set.  We
state the following lemma required in the proof of the rank guarantees
and prove it in the appendix.
\begin{lemma}
\label{lemma:detPnonzero}
Consider a matrix $\mathbf{P}\in \ext^{R\times R}$ defined as,
\begin{align*}
\mathbf{P}
&=
\left[
\begin{array}{c}
\g{1}\\
\g{2} \\
\vdots\\
\g{R}
\end{array}
\right]
\left[
\begin{array}{cccc}
1 & \ldots & 1 & 1 \\
\xi^{2^{R-1}} & \ldots & \xi^{2} & \xi \\
(\xi^{ 2 })^{2^{R-1} } & \ldots & (\xi^{2})^2 & \xi^{2} \\
\vdots & & \vdots & \\
(\xi^{(M_t-1)} )^{2^{R-1} } & \ldots & (\xi^{(M_t-1)})^2 & \xi^{(M_t-1)} \\
\end{array}
\right]
\end{align*}
where $\xi=\alpha^{(2^{R}-1)(\nu+1)}$ and the vectors
$\g{1},\ldots,\g{R} \in \Gamma^{M_t\times 1}$ are linearly independent
over $\ext$. If,
\begin{align*}
T & \geq  (2^R-1)\nu +  (2^R-1)(\nu+1) \left( (M_t-2) (2^{R}-1)+R \right).
\end{align*}
then $det(\mathbf{P})\neq 0$.
\end{lemma}

\begin{theorem}
Let $f(x) = \sum_{l=0}^{R-1}f_l x^{2^{l}}$ as in (\ref{eq:Fdef}) and
$T\geq T_{thr}$. Then for $\mathcal{S}$ defined in (\ref{eq:Sdef}),
$\frac{1}{T}\log|\mathcal{S}|\geq R-\frac{\nu M_t}{T} $ and $\forall
f\in\mathcal{S},
\mbox{rank}(\Ubf_f)\geq (M_t-R+1)(\nu+1)$ over the binary field.
\label{rate2}
\end{theorem}
\begin{proof}
The rate bound is directly from Theorem \ref{theorem_rate}.  If
$\mathcal{O}=\{\Ubf_f:f\in \mathcal{S}\}$ has rank distance
$(\nu+1)M_t- \mathcal{D}$ then there exists a vector $\ubf_{f}\neq \mathbf{0}$
for some $f \in \mathcal{S}$ such that the corresponding binary matrix
$\mathbf{U}_{f}$ has binary rank equal to $(\nu+1)M_t-\mathcal{D}$ (as the code
is linear). Equivalently there exists some $f \in \mathcal{S}$ for
which there exists a binary vector space $\mathcal{B}_f \subseteq
\mathbb{F}_2^{(\nu+1)M_t}$ of dimension $\mathcal{D}$ such that for every
$\mathbf{b} \in \mathcal{B}_f $, just as we saw in
(\ref{eq:NullSpUfSmallUfEq}), we have
\begin{align}
\mathbf{b}^t \mathbf{U}_{f} & = \mathbf{0}  \iff  \mathbf{b}^t \mathbf{u}_{f}  = 0
\end{align}
Note that the size of $\mathcal{B}_f$ is $2^\mathcal{D}$. Rewriting the above we
have that $ \forall \mathbf{b} \in \base^{1\times M_t(\nu+1)}$ and
$\mathbf{b}\in\mathcal{B}_f$,
\begin{align}
\label{eqn:intermsofb}
\underbrace{
\left[
\begin{array}{cccc}
b_1 & b_2 & \ldots & b_{(\nu+1)M_t}
\end{array}
\right] 
}_\mathbf{b}
\left[\begin{array}{c} 
f(1)\\
f(\xi)\\
\vdots\\
f(\xi^{(M_t-1)})\\
\alpha f(1)\\
\vdots \\
\alpha^{\nu} f(\xi^{(M_t-1)})
\end{array}
\right] & = 0
\end{align}
Let the function $\Psi$ be as in (\ref{eqn:DefOmega}) such that it
maps $\mathcal{B}_f$ to $\mathcal{G}_f$. Note that, since $\Psi$ is a
one-to-one mapping, as seen in (\ref{eqn:DefOmega}) in Section
\ref{subsec:def}, we immediately see that
\begin{align*}
|\mathcal{B}_f| & = |\mathcal{G}_f|.
\end{align*}
With the representation $\mathbf{g}=\Psi(\mathbf{b})$,
(\ref{eqn:intermsofb}) can be rewritten as,
\begin{align*}
\underbrace{
\left[
\begin{array}{cccc}
g_1 & g_2 & \ldots & g_{M_t}
\end{array}
\right] 
}_\mathbf{g}
\underbrace{
\left[\begin{array}{c} 
f(1)\\
f(\xi)\\
\vdots\\
f(\xi^{(M_t-1)})
\end{array}
\right]
}_{\mathbf{c}_f}
 & = 0
\end{align*}
where $g_i \in \Gamma$, or equivalently as
\begin{align*}
\left[
\begin{array}{cccc}
g_1 & g_2 & \ldots & g_{M_t}
\end{array}
\right] 
\underbrace{
\left[
\begin{array}{cccc}
1 & \ldots & 1 & 1 \\
\xi^{2^{R-1}} & \ldots & \xi^{2} & \xi \\
(\xi^{ 2 })^{2^{R-1} } & \ldots & (\xi^{2})^2 & \xi^{2} \\
\vdots & & \vdots & \\
(\xi^{(M_t-1)} )^{2^{R-1} } & \ldots & (\xi^{(M_t-1)})^2 & \xi^{(M_t-1)} \\
\end{array}
\right]
}_{\mathbf{W} \in \ext^{M_t\times R}}
\left[\begin{array}{c} 
f_{R-1}\\
f_{R-2}\\
\vdots\\
f_0
\end{array}
\right] & = 0
\end{align*}
If the only element in $\mathcal{G}_f$ is the all zero vector then,
$\mathcal{D}=0$, $\mathbf{U}_f$ has full binary rank, we have already shown the
result in Theorem \ref{rate1}. If not, by theorem \ref{theorem:basis}
there exists a set of minimal vectors,
$\mathcal{M}=\{\g{1},\g{2},\ldots,\g{d}\}$ for $\mathcal{G}_f$.

If $d\leq R-1$ it implies that $|\mathcal{G}_f|\leq 2^{(R-1)(\nu+1)}$,
and therefore $\mathcal{D}=dim(\mathcal{B}_f)\leq (R-1)(\nu+1)$ which
in turn would imply that all matrices in $\mathcal{O}$ have rank at
least $(M_t-R+1)(\nu+1)$. We will prove that $d \leq R-1$ by
contradiction. Let us assume that there are more than $R-1$ such
minimal vectors {\em i.e.,} $d>R-1$. Taking any $R$ of the minimal
vectors of the solution space $\mathcal{G}_f$ we conclude that,
\begin{align}
\label{eqn:defP}
\underbrace{
\left[
\begin{array}{c}
\g{1}\\
\g{2} \\
\vdots\\
\g{R}
\end{array}
\right]
\left[
\begin{array}{cccc}
1 & \ldots & 1 & 1 \\
\xi^{2^{R-1}} & \ldots & \xi^{2} & \xi \\
(\xi^{ 2 })^{2^{R-1} } & \ldots & (\xi^{2})^2 & \xi^{2} \\
\vdots & & \vdots & \\
(\xi^{(M_t-1)} )^{2^{R-1} } & \ldots & (\xi^{(M_t-1)})^2 & \xi^{(M_t-1)} \\
\end{array}
\right]
}_{\mathbf{P}}
\left[ 
\begin{array}{c}
f_{R-1} \\
\vdots \\
 f_1 \\
f_0
\end{array}
\right]
& = \mathbf{0}_{R \times 1}
\end{align}
where $\mathbf{P}\in\ext^{R\times R}$.  This is possible iff,
\begin{align*}
det (\mathbf{P})&=0
\end{align*}
As shown in lemma \ref{lemma:detPnonzero} by the linear independence
of $\{\alpha^0,\alpha^1,\ldots,\alpha^{T-1}\}$ it follows that the
determinant can never be zero. Therefore there can be at most $R-1$
basis vectors and from (\ref{eqn:sizeofD}) and property \ref{prop3} of
theorem \ref{theorem:basis} since,
\begin{align*}
|\mathcal{B}_f|& = 2^\mathcal{D} \\
|\mathcal{G}_f|& \leq 2^{(R-1)(\nu+1)}\\
|\mathcal{B}_f|& =|\mathcal{G}_f|.
\end{align*}
we conclude that $\mathcal{D}\leq(R-1)(\nu+1)$. Therefore all matrices in
$\mathcal{O}$ have rank at least $(M_t-R+1)(\nu+1)$.
\qed
\end{proof}

The consequence of Theorem \ref{rate2} is that
$\mathcal{K}_{\nu,d}=\left\{ \mathbf{C}_{f} : f
\in \mathcal{S}\right\}$ satisfies the requirements of definition 
\ref{suitable_isi} and therefore can be used to construct
diversity embedded codes for fading ISI channels as done in Theorem
\ref{thm:MLC_ISI}.



\section{Examples and Discussion}
\label{sec:examples}
We will start off by giving an example of a code which has full
diversity equal to $M_t$ when transmitted over the flat fading channel
but does not have the maximum possible diversity of $(\nu+1)M_t$ when
transmitted over an ISI channel with $\nu$ taps. 

{\em Example 1:} Consider construction of a code for $M_t=2$, $T=5$
with rate $R=1$ and BPSK signaling using code constructions given in
\cite{Gabidulin,LuKumar03}. To design these
codes, use the field extension $\mathbb{F}_{\field^5}$ with the
primitive polynomial given by $x^5+x^4+x^2+x+1$ and the primitive
element $\alpha$. Define,
\begin{align*}
f(x) & = f_0 x
\end{align*}
where $f_0 \in \mathbb{F}_{field^5}$ depends on the input message. The
space time codeword is obtained as,
\begin{align}
\mathbf{C}_{f_0} & = \left[\begin{array}{cc} \mathbf{f}^t(1) & \mathbf{f}^t(\alpha) \end{array} \right]^t
\end{align}
where $\mathbf{f}(\alpha^i)$ is the representation of $f(\alpha^i)$ as
a binary $1 \times 5 $ row vector and$ \mathbf{C}_{f} \in \base^{2
\times 5 }$. As was shown in \cite{Gabidulin,LuKumar03} this code
achieves full diversity $M_t=2$ {\em i.e.}, $\mathbf{C}_{f_0}$ has
rank $2$ for all nonzero $f_0\in \mathbf{F}_{\field^5}$.

Now assume that we use this code for transmission over an ISI channel
with $\nu=1$. Since this is a linear code, the rank distance of the
code is the minimum rank of a nonzero codeword. Therefore the space
time codeword corresponding to $f_0=1$ is given by,
\begin{align}
\mathbf{C}_{1} & =
\left[ \begin{array}{ccccc}
1 & 0 & 0 & 0 & 0 \\
0 & 1 & 0 & 0 & 0 
\end{array}
\right].
\end{align}
When transmitted over the ISI channel we see that the equivalent space
time codeword is given by,
\begin{align}
\Theta\left( \mathbf{C}_{1}\right) & =
\left[ \begin{array}{ccccc}
1 & 0 & 0 & 0 & 0 \\
0 & 1 & 0 & 0 & 0 \\
0 & 1 & 0 & 0 & 0 \\
0 & 0 & 1 & 0 & 0 
\end{array}
\right].
\end{align}
Clearly since,
\begin{align*}
\mbox{rank}(\Theta\left( \mathbf{C}_{1}\right))& = 3  < 4
\end{align*}
we conclude that the space time codeword which achieves full diversity
$M_t=2$ over the flat fading channel does not achieve the maximum
possible diversity of $(\nu+1)M_t=4$ over the ISI channel.

{\em Example 2:} Similarly this can be shown to hold true for any
diversity point. Consider for example the case of $M_t=3$, $T=7$,
$R=2$ and BPSK signaling using code constructions given in
\cite{Gabidulin,LuKumar03}.  Use the field
extension $\mathbb{F}_{\field^7}$ with the primitive element
$\alpha$. Define,
\begin{align*}
f(x) & = f_1 x^2 +f_0 x
\end{align*}
where $f_0 \in \mathbb{F}_{\field^7}$ depends on the input message as
before. The space time codeword is obtained as,
\begin{align}
\mathbf{C}_{f_0} & = \left[\begin{array}{ccc} \mathbf{f}^t(1) & \mathbf{f}^t(\alpha) & \mathbf{f}^t(\alpha^2) \end{array} \right]^t
\end{align}
where $\mathbf{f}(\alpha^i)$ is the representation of $f(\alpha^i)$ as
a binary $1 \times 5 $ row vector and$ \mathbf{C}_{f} \in \base^{2
\times 5 }$. As was shown in \cite{Gabidulin,LuKumar03} this code
achieves diversity $d=2$ {\em i.e.}, $\mathbf{C}_{f_0}$ has rank $2$
for all nonzero $f_0\in \mathbf{F}_{\field^7}$. But it can be seen as
before that the space time codeword corresponding to $(f_1,f_0)=(0,1)$
does not achieve the maximum possible diversity of $(\nu+1)M_t$ when
transmitting over the ISI channel with $\nu$ taps.

{\em Example 3:} Consider construction of a BPSK code for $M_t=2$,
$\nu=1$, $T=5$ with rate $R=1$ and hence $R^{eff}=\frac{3}{5} $. To
design these codes, use the field extension $\mathbb{F}_{\field^5}$
with the primitive polynomial given by $x^5+x^4+x^2+x+1$ and the
primitive element $\alpha$. The set of codeword polynomials which
satisfy the constraints in (\ref{eq:Sdef}) are given by,
\begin{align}
\mathcal{S} & =
\left\{0,\alpha,\alpha^{17},\alpha^{19},\alpha^{21},\alpha^{24},\alpha^{26},\alpha^{31} \right\}.
\end{align}
This set is of cardinality
\begin{align*}
|\mathcal{S}| & = 2^{RT-\nu M_t} = 2^{5-2} =8.
\end{align*}
Corresponding to every element $f$ in $\mathcal{S}$ consider the
codeword vector,
\begin{align}
\mathbf{c}_{f} & = \left[\begin{array}{cc} f(1) & f(\alpha^{2}) \end{array} \right]^t
\end{align}
where $\mathbf{c}_{f} \in \mathbb{F}_{\field^5}^{2 \times 1}$. Let
$\mathbf{C}_{f} \in \base^{2 \times 5}$ be the representation of
each element of $\mathbf{c}_{f}$ in the basis
$\{\alpha^0,\alpha^1,\alpha^{2},\alpha^{3},\alpha^{4} \}$, {\em i.e.},
\begin{align}
\mathbf{C}_{f} & = \left[\begin{array}{cc} \mathbf{f}^t(1) & \mathbf{f}^t(\alpha^{2}) \end{array} \right]^t
\end{align}
where $\mathbf{f}(\alpha^i)$ is the representation of
$f(\alpha^i)$ as a binary $1 \times 5 $ row vector. Then the $2
\times 5 $ space time code has rate,
\begin{align*}
R & = 1-\frac{\nu M_t}{ T}  = 1 -\frac{2}{5} = \frac{3}{5}
\end{align*}
and gives diversity $4$ when transmitted over the ISI channel with
$\nu=1$. The corresponding $8$ codewords $\Xbf^{(1)}$ as given in
(\ref{eq:X1def}) are,
\begin{align*}
\left[
\begin{array}{ccccc}
 0 & 0 & 0 & 0 & 0 \\
 0 & 0 & 0 & 0 & 0
\end{array}
\right],
& 
\left[
\begin{array}{ccccc}
1  & 0  & 0  & 0 & 0 \\
0  & 0  & 1  & 0 & 0
\end{array}
\right], 
\\
\left[
\begin{array}{ccccc}
 0 & 1 & 0 & 0 & 0 \\
 0 & 0 & 0 & 1 & 0
\end{array}
\right],
& 
\left[
\begin{array}{ccccc}
 1 & 1 & 0 & 0 & 0 \\
 0 & 0 & 1 & 1 & 0
\end{array}
\right] 
\\
\left[
\begin{array}{ccccc}
0  & 0 & 1 & 1 & 0 \\
1  & 1 & 1 & 0 & 0
\end{array}
\right],
& 
\left[
\begin{array}{ccccc}
 1 & 0 & 1 & 1 & 0 \\
 1 & 1 & 0 & 0 & 0
\end{array}
\right],
\\ 
\left[
\begin{array}{ccccc}
 0 & 1 & 1 & 1 & 0 \\
 1 & 1 & 1 & 1 & 0
\end{array}
\right],
& 
\left[
\begin{array}{ccccc}
 1 & 1 & 1 & 1 & 0 \\
 1 & 1 & 0 & 1 & 0
\end{array}
\right].  
\end{align*}

\begin{figure}
\begin{center}
\includegraphics{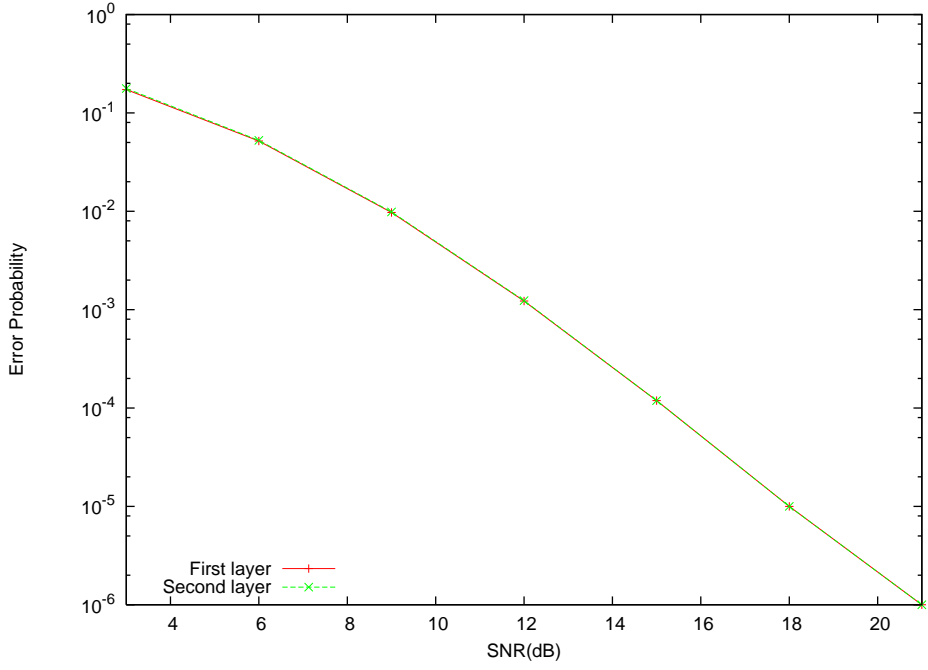}
\end{center}
\caption{Error Performance of full diversity codes with $M_t=2$, $\nu=1$, $R_1=R_2=\frac{3}{5}$ and $d_1=d_2=4$.}
\label{fig:Pe_full_diversity}
\end{figure}

In figure \ref{fig:Pe_full_diversity} we give the performance of a full
diversity code which is designed for $M_t=2$, $M_r=1$, $\nu=1$ and 4-QAM signal
constellation. We plot the logarithm of the error probability as a
function of SNR (in dB). Note that the slope of the error probability
curve is approximately equal to $4$ which is expected since we are
using full diversity codes on both the layers. 

From the construction of these codes, one might be tempted to conclude
that the analysis for these codes is quite similar to that of cyclic
codes. But the peculiar structure of the solution space {\em i.e.} the
fact that given a vector in the solution space $\mathcal{B}$, not all
circular shifts of the vector remain in $\mathcal{B}$, makes it
difficult to analyze. The main contribution of this work is the
construction of binary matrices with a particular structure which
consequently characterizes the rate diversity tradeoff for the ISI
channel. Note that as seen in Example 1 and 2 codes which give guaranteed
diversity orders for flat fading MIMO channel, when used for
transmission over ISI channel do not necessarily give the
multiplicative diversity gain of $(\nu+1)$. The tools and techniques
developed over here could also have independent interest in designing
codes in various other wireless or distributed settings.

\section*{Acknowledgments}
We would like to thank Amin Shokrollahi for interesting and helpful
discussions about this work, and in particular for discussions on the
proof of Theorem \ref{theorem_rate}.

\section{Appendix}

\begin{proof} [of Lemma \ref{lemma:Prop3_Generalized}] We have that 
for some subset $S \subseteq\{1,\ldots,p\}$ there exist $\left\{
\gamma_i\right \}_{i\in S}$ such that,
\begin{align}
\label{eqn:MaxDegree}
deg(\sum_{i\in S} \gamma_i \g{i}) & < \max_{i \in S} deg(\gamma_i
\g{i})
\end{align}
Let $t=\max_{i \in S} deg(\gamma_i \g{i})$ and define
$\mathcal{T}=\left\{ i: deg(\gamma_i\g{i})=t \right \}$. 
Note that we have,
\begin{equation}
\label{eq:DegEquality}
deg(\gamma_i\g{i}) = deg(\alpha^{deg(\gamma_i)} \g{i})
\end{equation}
This allows us to see that (\ref{eqn:MaxDegree}) implies that,
\begin{align}
\label{eqn:MaxDegree2}
deg\left(\sum_{i\in \mathcal{T} } \alpha^{deg(\gamma_i)} \g{i}\right) & < t.
\end{align}
Denote $w=\min_{i\in \mathcal{T}}(deg(\gamma_i))$ to be the minimum
degree of $\gamma_i$ for $i\in \mathcal{T}$ and $k=\argmin_{i\in
\mathcal{T}}(deg(\gamma_i))$.  Define $S'=\mathcal{T}\backslash \{k\}$
where $\backslash$ is the set difference operator.

If $w=0$ then we have,
\begin{align*}
deg(\g{k}+ \sum_{i\in S' } \alpha^{deg(\gamma_i)} \g{i}) & < t=deg (\g{k}) 
\end{align*}
and,
\begin{align*}
t & =deg\left( \alpha^{deg(\gamma_i)} \mathbf{g}^{(i)} \right) \leq
deg\left(\mathbf{g}^{(k)}\right)=t \quad \forall i \in S'
\end{align*}
which shows that if $w=0$, then the claim is true.

If $w\neq 0$ we can take out the common $\alpha^w$ factor of
$\sum_{i\in \mathcal{T} } \alpha^{deg(\gamma_i)} \g{i}$. Then we have,
\begin{align*}
deg(\g{k}+ \sum_{i\in S' } \alpha^{deg(\gamma_i)-w} \g{i}) & < (t-w) =deg (\g{k})
\end{align*}
and,
\begin{align*}
t-w & =deg\left( \alpha^{deg(\gamma_i)-w} \mathbf{g}^{(i)} \right)
\leq deg\left(\mathbf{g}^{(k)}\right)=t-w \quad \forall i \in S'
\end{align*}
Hence the claim is proved.
\qed
\end{proof}

To prove the Lemma \ref{lemma:detPnonzero} we will make use of the
Cauchy Binet formula reproduced here for completeness.

\begin{definition}{\bf Cauchy Binet Formula} \cite{horn}
Let $\mathbf{A}$ be a $m\times n$ matrix and $\mathbf{B}$ be a
$n\times m$ matrix. If $S$ is a subset of $\{1,\ldots, n \}$ with $m$
elements, let $\mathbf{A}_S$ represent the $m\times m$ matrix whose
columns are those columns of $\mathbf{A}$ that have indices from
$S$. Similarly, let $\mathbf{B}_S$ represent the $m\times m$ matrix
whose rows are those rows of $\mathbf{B}$ that have indices from
$S$. The Cauchy-Binet formula then states that,
\begin{align}
det(\mathbf{AB}) & =  \sum_{S} det(\mathbf{A}_S) \det(\mathbf{B}_S)
\end{align}
where the sum extends over all possible subsets $S$ of $\{ 1,\ldots,
n\}$ with $m$ elements.
\end{definition}
Note that the Cauchy Binet formula holds for matrices with entries
from any commutative rings. Given this definition, the proof of lemma
\ref{lemma:detPnonzero} proceeds as follows.

\begin{proof} [of Lemma \ref{lemma:detPnonzero}]
The matrix $\mathbf{P}$ is given by,
\begin{align*}
\mathbf{P} & = \underbrace{
\left[
\begin{array}{c}
\g{1}\\
\g{2} \\
\vdots\\
\g{R}
\end{array}
\right]
}_{\Mbf\in\Gamma^{R \times M_t}}
\underbrace{
\left[
\begin{array}{cccc}
1 & \ldots & 1 & 1 \\
\xi^{2^{R-1}} & \ldots & \xi^{2} & \xi \\
(\xi^{ 2 })^{2^{R-1} } & \ldots & (\xi^{2})^2 & \xi^{2} \\
\vdots & & \vdots & \\
(\xi^{(M_t-1)} )^{2^{R-1} } & \ldots & (\xi^{(M_t-1)})^2 & \xi^{(M_t-1)} \\
\end{array}
\right]
}_{\mathbf{W} \in \ext^{M_t\times R}}
\end{align*}
Using Gaussian elimination (which can be applied over any finite
field), we reduce the matrix $\Mbf$ to its row echelon form,
\begin{align*}
\mathbf{\breve{Y}} & = 
\left[
\begin{array}{c}
\bg{1}\\
\bg{2} \\
\vdots\\
\bg{R}
\end{array}
\right]
\end{align*}
where 
\begin{align*}
deg(\bg{k}) & \leq  2^{(R-k)}\nu.
\end{align*}
Note that this pivoting and reduction to a row echelon form is a full
rank operation and preserves the rank of $\mathbf{P}$. Therefore, 
\begin{align*}
det(\mathbf{P})&= K det \left( \mathbf{\breve{Y}} \mathbf{W}\right)
\end{align*}
where $K\in \ext$ and $K\neq 0$. Let the columns containing the pivots
 in $\mathbf{\breve{Y}}$ be denoted by $\tilde{S}$. Therefore by the
 Cauchy Binet formula, we have
\begin{align}
\label{eqn:detP}
K^{-1} det(\mathbf{P})&= det \left( \mathbf{\breve{Y}}_{\tilde{S}} \right) det \left( \mathbf{W}_{\tilde{S}} \right)+ \sum_{S\neq \tilde{S} } det \left( \mathbf{\breve{Y}}_S \right) det \left( \mathbf{W}_S \right)
\end{align}
Note that for all $S$ such that $det\left(\Mbf_S\right)\neq 0$
the maximum coefficient of $\xi$ in $det\left(\mathbf{W}_S \right)$ is
less than the maximum coefficient of $\xi$ in
$det\left(\mathbf{W}_{\tilde{S}}\right)$ by at least $1$. Therefore,
\begin{align*}
deg(det\left(\mathbf{W}_{\tilde{S}}\right)) -deg(det\left(\mathbf{W}_S \right)) & \geq (2^R-1)(\nu+1)
\end{align*}
Also note that,
\begin{align*}
deg(det(\Mbf_S))-deg(det(\Mbf_{\tilde{S}})) & \leq \nu +2 \nu +2^2 \nu+\ldots + 2^{R-1}\nu \\
& = \nu (2^R-1)
\end{align*}
Therefore, 
\begin{align*}
deg(det\left(\mathbf{W}_{\tilde{S}}\right) det(\Mbf_{\tilde{S}}) ) - deg(det\left(\mathbf{W}_S \right)det(\Mbf_S)) & \geq (2^R-1)(\nu+1) - \nu (2^R-1) \\
& > 0
\end{align*}
Therefore by the linear independence of
$\{1,\alpha,\ldots,\alpha^{T-1}\}$ we can conclude that there exists a
term in $det\left(\mathbf{W}_{\tilde{S}}\right)
det(\Mbf_{\tilde{S}})$ with a power of $\alpha$ which in not
canceled by any other term in the equation (\ref{eqn:detP}). Therefore
we conclude that $K^{-1} det(\mathbf{P})\neq 0$ implying
$det(\mathbf{P})\neq 0$. Hence proved.
\qed
\end{proof}

\small

\bibliographystyle{plain}

\normalsize

\end{document}